\documentclass[fleqn,usenatbib]{mnras}
\usepackage{newtxtext,newtxmath}
\usepackage[T1]{fontenc}
\DeclareRobustCommand{\VAN}[3]{#2}
\let\VANthebibliography\thebibliography
\def\thebibliography{\DeclareRobustCommand{\VAN}[3]{##3}\VANthebibliography}
\usepackage{amsmath}
\usepackage{bm}
\usepackage{ascmac}
\usepackage{chngpage}
\usepackage{graphicx}
\usepackage{lscape}
\usepackage{pdflscape}
\usepackage{color}
\usepackage{comment}
\defcitealias{10.1093/mnras/stac1289}{Paper~I}

\title[X-ray plasma flow and turbulence in WR140]{X-ray plasma flow and turbulence in the colliding winds of WR140}

\author[A. Miyamoto et al.]{
Asca Miyamoto,$^{1}$\thanks{E-mail: miyamoto-asuka@ed.tmu.ac.jp (AM)}
Yasuharu Sugawara,$^{2}$
Yoshitomo Maeda,$^{2}$
Manabu Ishida,$^{1,2}$
Kenji Hamaguchi,$^{3,4}$\vspace{1ex}\\
{\rm\Large Michael Corcoran,$^{3}$
Christopher M. P. Russell,$^{5}$
and Anthony F. J. Moffat$^{6}$}.\\
$^{1}$Department of Physics, Tokyo Metropolitan University, 1-1 Minami-Osawa, Tokyo 192-0397,
Japan\\
$^{2}$The Institute of Space and Astronautical Science/JAXA, 3-1-1 Yoshinodai, Chuo-ward,
Sagamihara, Kanagawa 252-5210, Japan\\
$^{3}$CRESST II and X-ray Astrophysics Laboratory, NASA/GSFC, Greenbelt, MD 20771, USA\\
$^{4}$Department of Physics, University of Maryland, Baltimore County, 1000 Hilltop Circle, Baltimore, MD 21250, USA\\
$^{5}$Department of Physics and Astronomy, Bartol Research Institute, University of Delaware, Newark, DE 19716, USA\\
$^{6}$D\'epartement de physique and Centre de Recherche en Astrophysique du Qu\'ebec (CRAQ), Universit\'e de Montr\'eal, C.P. 6128, Succ. A, Canada
}

\date{Accepted XXX. Received YYY; in original form ZZZ}

\pubyear{2023}

\begin{document}
\label{firstpage}
\pagerange{\pageref{firstpage}--\pageref{lastpage}}
\maketitle

\begin{abstract} 
We analyse {\it XMM-Newton} RGS spectra of Wolf-Rayet (WR) 140, an archetype long-period eccentric WR+O colliding wind binary. 
We evaluate the spectra of O and Fe emission lines and find that the plasmas emitting these lines have the largest approaching velocities with the largest velocity dispersions
between phases 0.935 and 0.968 where the inferior conjunction of the O star occurs.
This behaviour is the same as
that of the Ne line-emission plasma presented in our previous paper. 
We perform diagnosis of electron number density $n_{\rm e}$ using He-like triplet lines of O and Ne-like Fe-L lines. 
The former results in a conservative upper limit of $n_{\rm e} \lesssim 10^{10}$-10$^{12}$ cm$^{-3}$ on the O line-emission site, while the latter can not impose any constraint on the Fe line-emission site because of statistical limitations. 
We calculate the line-of-sight velocity and its dispersion separately along the shock cone. By comparing the observed and calculated line-of-sight velocities,
we update the distance of the Ne line-emission site from the stagnation point.
By assuming radiative cooling of the Ne line-emission plasma using the observed temperature and the local stellar wind density, we estimate the line-emission site extends along the shock cone by at most $\pm$58 per cent (phase 0.816) of the distance from the stagnation point.
In this framework, excess of the observed velocity dispersion over the calculated one is ascribed to
turbulence in the hot-shocked plasma at earlier orbital phases of 0.816, 0.912, and 0.935, with the largest velocity dispersion of 340-630 km~s$^{-1}$ at phase 0.912.
\end{abstract}


\section{Introduction}
A classical Wolf-Rayet (cWR) star is the final stage in the evolution of a massive star. It generally has a surface temperature of $>$30,000 K, luminosity of $\sim10^6$ L$_\odot$, and large initial mass of $>$25 M$_\odot$. A cWR star emits high-velocity stellar wind with a terminal speed of approximately 2000~km~s$^{-1}$ and large mass-loss rate of $> 10^{-5}$~M$_\odot$~yr$^{-1}$, producing a spectrum with broad emission lines. 
cWR stars are further classified into three broad subtypes according to their spectral characteristics: WN (primarily He and N emission lines), WC (no N and primarily He and C), and WO (O and WC emission lines). These subtypes are divided into subclasses according to their degree of ionisation. 
cWR stars explode as core-collapse supernovae, 
 wherein the WN and WC stars become H-poor type-SN Ib and WO stars become type-SN Ic, the latter owing to the absence of an outer He layer.

\citet{1976SvA....20....2P} and \citet{1976SvAL....2..138C} first studied the production of X-ray emissions by the collision of dense stellar winds in massive binary stars \citep[see also][]{1978Natur.273..645C}. 
They showed that the gas temperature reached 
$\sim$10$^7$-10$^8$~K with X-ray luminosities of $10^{33}$-$10^{35}$ erg~s$^{-1}$. Initial
X-ray observations 
\citep{1979ApJ...234L..55S, 1982IAUS...99..577M, 1985Natur.313..376C} showed that WR
stars emit X-rays, irrespective of whether they belong to binaries, as described by
\citet{1976SvA....20....2P}. \citet{1987ApJ...320..283P} conducted a uniform analysis of the 48
WR stars, which were observed with the {\it Einstein} X-ray Observatory. Their luminosities in the soft band
(0.1-4~keV) were in the range from $10^{32}$ to $10^{34}$ erg~s$^{-1}$. By incorporating the radio data,
\citet{1987ApJ...320..283P} concluded that X-rays from the brightest 
group of the {\it Einstein} samples originated directly from
the colliding stellar winds, as predicted by \citet{1976SvA....20....2P}, or from the Compton scattering
of photosheric radiation by relativistic electrons accelerated by surface magnetic fields of up to
a few hundred gauss, although \citet{2014ApJ...781...73D} claimed that no significant global magnetic field existed.

WR140 (HD193793), the target of this study, is a WR+ O binary composed of WC7pd and O5.5fc stars \citep{2011MNRAS.418....2F}, orbiting each other with a period of just under 8 years. Both stars expel high-velocity stellar winds, and their collision creates shocks that heat and compress the hot plasma, which then emits X-rays. 
WR140 is among the brightest massive binaries observed by {\it Einstein} and 
has been detected since the earliest stages of X-ray astronomy by {\it Uhuru}
\citep{1978ApJS...38..357F}, {\it HEAO-1} \citep{1984ApJS...56..507W} and {\it EXOSAT} \citep{1990MNRAS.243..662W}.

Detailed X-ray spectrometry became possible with later X-ray astronomy satellites
\citep{1990PASJ...42L...1K, 1994PASJ...46L..93K, 2000ApJ...538..808Z, 2005ApJ...629..482P, 2011BSRSL..80..653D, 2015PASJ...67..121S, 2021ApJ...923..191P}. \citet{1990PASJ...42L...1K}
measured the X-ray spectrum of WR140 using the {\it Ginga} observatory. They determined the X-ray flux of 2-6~keV to be $1.5\times 10^{-11}$ erg~s$^{-1}$~cm$^{-2}$, which results in a luminosity of
$4.1\times 10^{33}$ erg~s$^{-1}$ by assuming a distance of 1518~pc
\citep{2021MNRAS.504.5221T}. \citet{1994PASJ...46L..93K} observed WR140 using {\it ASCA} at the phase
when the WR star was nearly in front of the O star. They found that the X-ray spectrum was heavily
absorbed by $N_{\rm H} \simeq 3\times 10^{22}$~cm$^{-2}$. A series of X-ray observations of up to
$\sim$10~keV across the periastron passage were performed using {\it XMM-Newton}
\citep{2011BSRSL..80..653D} and {\it Suzaku} \citep{2015PASJ...67..121S}. The researchers detected an increase in
line-of-sight absorption as the 
stars approached the periastron
passage. \citet{2015PASJ...67..121S} measured maximum plasma temperatures of 
3.0-3.5~keV (35-41 MK) over a phase interval of 2.904-3.000.

In our first study \citep[][hereafter referred to as Paper I]{10.1093/mnras/stac1289}, we analysed the data from WR140 observed using the reflection grating spectrometer \citep[RGS; ][]{2001A&A...365L...7D} onboard {\it XMM-Newton} \citep{2001A&A...365L...1J} over a period of 8 years and measured the plasma temperature, line-of-sight velocity, and velocity dispersion of the Ne emission lines at different orbital phases. We calculated the shape of the shock cone based on the balance of ram pressure between the stellar winds and evaluated the location of the Ne line-emission site on the shock cone by comparing the ratio of the expected line-of-sight velocity to the expected velocity dispersion with that of the observed value. 
We also constrained the electron number densities using the intensity ratio of He-like triplets of Ne at different orbital phases.

In this study, we aim to advance the understanding of the nature of shock cone plasma in WR140. 
The remainder of this paper is organised as follows: In Section 2, we describe the data used in this study and the data reduction method. In Section 3, we explain the data analysis methods adopted to derive the line-of-sight velocities and redtheir dispersions of O and Fe emission lines. These emission lines show a similar velocity trend to that of the Ne emission lines presented in \citetalias{10.1093/mnras/stac1289}. We also attempt to constrain the densities using the intensity ratios of these line components. In Section 4, we calculate the line-of-sight velocity and its dispersion (separately, not their ratio) of the plasma flowing in the shock cone, whose geometry was obtained in \citetalias{10.1093/mnras/stac1289}, from which the location of the Ne line-emission site is updated from \citetalias{10.1093/mnras/stac1289}. Additionally, the location of the O line-emission site is determined. We evaluate spatial extent of the Ne and O line-emission sites along the shock cone. For the first time, we report that excess of the observed line-of-sight velocity dispersion can be explained by turbulence in the X-ray plasma flow, using spatial extent of the Ne line-emission site evaluated from the temperature of the Ne line-emission plasma and its cooling time.
Finally, in Section 5, we summarise the results of this study.

In this paper, all errors quoted are at the 90 per cent confidence level unless otherwise mentioned.

\section{Observations and data reduction}
\subsection{Observations}
We analyse 10 datasets obtained at different orbital phases using the RGS \citep{2001A&A...365L...7D} onboard {\it XMM-Newton} \citep{2001A&A...365L...1J}, covering a period of just over 8 years, from May 2008 to June 2016. The data all had individual exposure times of
more than 18 ks. 
The orbital parameters adopted in this study are those used in \citet{2011ApJ...742L...1M} and are summarised in Table 1 of \citetalias{10.1093/mnras/stac1289}.
The most recent parameters are provided in \citet{2021MNRAS.504.5221T}; however, for consistency with \citetalias{10.1093/mnras/stac1289}, we continue to employ 
those of \citet{2011ApJ...742L...1M} in this paper.
The apparent binary orbit projected onto the celestial sphere is sketched in \citet{2011ApJ...742L...1M}.
The observation logs are summarised in Table 2 of \citetalias{10.1093/mnras/stac1289}.

\subsection{Data reduction}
As explained in \citetalias{10.1093/mnras/stac1289}, we extract the spectra of the first and second orders of RGS1 and RGS2 from the event files and create response files according to the standard data reduction method
using the HEAsoft (version 6.27.2\footnote{https://heasarc.gsfc.nasa.gov/docs/software/lheasoft/}) program provided by NASA’s GSFC and the SAS (version 19.1.0\footnote{https://www.cosmos.esa.int/web/xmm-newton/sas}) provided by ESA.
Figure 2 in \citetalias{10.1093/mnras/stac1289} shows the positions of the O star relative to the WR star, where the 10 RGS observations are made together with their RGS spectra. As \citetalias{10.1093/mnras/stac1289}, we analyse only the datasets at the orbital phases K (0.816), A (0.912), L (0.935), B (0.968), and D (0.987) where the O star is in front of the WR star, showing X-ray spectra with sufficient statistical quality. An inferior conjunction of the O star occurs between phases L (0.935) and B (0.968).

\section{Data Analysis}

\subsection{Line-of-sight velocity and its dispersion of O and Fe lines}
\subsubsection{\ion{O}{VII} and \ion{O}{VIII} K$\alpha$ lines}
Similar to the Ne lines in \citetalias{10.1093/mnras/stac1289}, we evaluate the line-of-sight velocity of K$\alpha$ lines of \ion{O}{vii,viii} and their dispersion. We perform spectral fitting by adopting a bvvapec*tbabs model using the energy bands of K$\alpha$ lines of \ion{O}{vii,viii} (0.55-0.59 keV and 0.635-0.670 keV).
The bvvapec model describes a velocity-broadened emission spectrum from an optically thin thermal plasma in collisional ionisation equilibrium, similar to the bvapec used in \citetalias{10.1093/mnras/stac1289}. Although bvvapec can change the abundance of elements with odd atomic numbers, all abundances that are not variable in bvapec are fixed at the solar abundances \citep{1989GeCoA..53..197A}. Consequently, the bvvapec model used in this study is identical to that used in \citetalias{10.1093/mnras/stac1289}, and we adopt the parameters shown in Table~4 
of \citetalias{10.1093/mnras/stac1289} as the best-fit parameters of full energy-band fits.

\begin{figure*}
	\begin{minipage}{0.498\textwidth}
    \includegraphics[width=\textwidth]{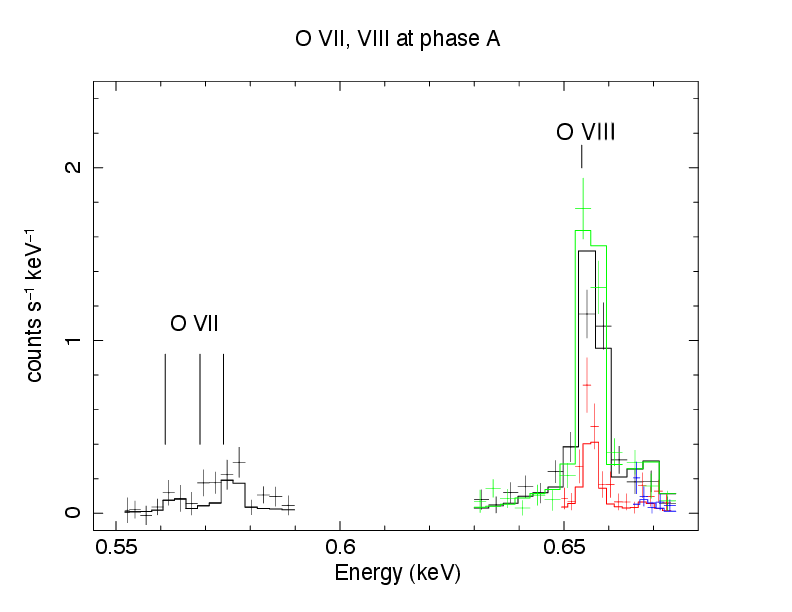}
    \end{minipage}
    \hfill
    \begin{minipage}{0.498\textwidth}
    \includegraphics[width=\textwidth]{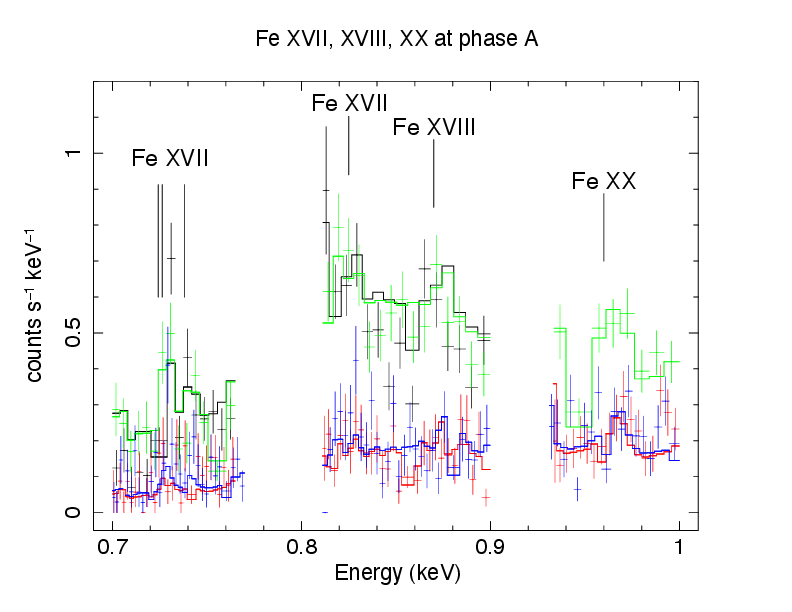}
    \end{minipage}
    \begin{minipage}{0.9\textwidth}
    \caption{Spectra of phase A (0.912) around the energy bins of \ion{O}{vii, viii} K$\alpha$ (left) and \ion{Fe}{xvii,xviii,xx} lines (right). Black and green colours are used for the first-order spectra of RGS1 and RGS2, respectively, while the red and blue colours are used for their second-order spectra, respectively. The best-fitting bvvapec*tbabs models are overlaid as histograms, whose parameters are summarised in Tables \ref{tab:O78bestfit} and \ref{tab:Feallbestfit}. 
    \label{fig:fitOFe}}   
    \end{minipage}
\end{figure*}

To evaluate the \ion{O}{vii,viii} lines,
we set the temperature, O abundance, line-of-sight velocity and its dispersion, emission measure of the bvvapec model, and hydrogen column density of the tbabs model as free parameters and fix the abundances other than O at the values obtained with the full energy-band fit \citepalias[Table 4 in][]{10.1093/mnras/stac1289}. 
The results are summarised in Table \ref{tab:O78bestfit} and Fig. \ref{fig:fitOFe} (left).
\begin{table*}
	\centering
	\caption{Best-fit parameters of the K$\alpha$ lines of \ion{O}{VII, VIII} with the bvvapec and tbabs models [see Fig.~\ref{fig:fitOFe} (left) for the plot].
    Hydrogen column density $N_{\rm H}$} and abundances other than O are fixed at the best-fit parameters in the full energy band \citepalias[Table 4 in][]{10.1093/mnras/stac1289}. The parameters $kT$, $Z_{\rm O}$, $v_{\rm los}$ (redshift of centroid energies of emission lines), $\sigma_{\rm obs}$ (broadening of emission lines), and the emission measure (EM) are allowed to vary.
	\label{tab:O78bestfit}
	\begin{tabular}{lcccccccccccr}
		\hline
		ID & K & A & L & B & D\\
		Phase &0.816&0.912&0.935&0.968&0.987\\
		\hline
		$kT$ (keV) & 0.233$^{+0.026}_{-0.031}$ & $0.210^{+0.033}_{-0.021}$ & 0.229$^{+0.070}_{-0.030}$ & $0.233^{+0.049}_{-0.036}$ & $0.215^{+0.0488}_{-0.024}$ \\
		$Z_{\rm O}$ & 0.259$^{+0.750}_{-0.103}$ & 0.602$^{+1.689}_{-0.370}$ & 0.122$^{+0.172}_{-0.040}$ & 0.080$^{+0.084}_{-0.018}$& 1.178$^{+8.855}_{-0.519}$\\
		$Z_{\rm Ne}$ & 0.775 (fixed) & 0.624 (fixed) & 0.598 (fixed) & 0.586 (fixed) & 1.191(fixed) \\
		$Z_{\rm Mg}$ & 0.133 (fixed) & 0.117 (fixed) & 0.130 (fixed) & 0.122 (fixed) & 0.316 (fixed)\\
		$Z_{\rm Si}$ & 0.131 (fixed) & 0.140 (fixed) & 0.146 (fixed) & 0.140 (fixed) & 0.664 (fixed)\\
		$Z_{\rm Fe}$=$Z_{\rm Ni}$ & 0.050 (fixed) & 0.041 (fixed) & 0.045 (fixed) & 0.046 (fixed) & 0.013 (fixed) \\
		$v_{\rm los}$ (km~s$^{-1}$) & $-1119^{+91}_{-82}$ & $-1088^{+81}_{-97}$ & $-1186^{+83}_{-75}$ & $-1180^{+84}_{-91}$ & $-716^{+170}_{-151}$ \\
		$\sigma_{\rm obs}$ (km~s$^{-1}$) & 719$^{+118}_{-93}$ & $537^{+117}_{-82}$ & 421$^{+159}_{-145}$ & $400^{+151}_{-105}$ & $821^{+206}_{-167}$ \\
		Norm. ($10^{-14}{\rm EM}/4\pi D^2$ cm$^{-5}$) & 0.144$^{+0.359}_{-0.055}$ & $0.250^{+0.647}_{-0.115}$ & 1.738$^{+10.646}_{-0575}$ & $2.074^{+13.046}_{-1.447}$ & $0.872^{+2.150}_{-0.500}$ \\
		$N_{\rm H}$ ($10^{22}$ cm$^{-2}$) & 0.676$^{+0.322}_{-0.104}$ & 0.848$^{+0.320}_{-0.086}$ & 0.951$^{+0.347}_{-0.058}$ & 0.944$^{+0.399}_{-0.132}$ & 1.383$^{+0.493}_{-0.149}$ \\
						\hline
		C-statistics (dof) & 446.16 (435) &499.01 (442) & 443.39 (441) & 475.06 (488)& 450.5 (433) &\\
		\hline
	\end{tabular}
\end{table*}
The line-of-sight velocity ranges from $-$700 to $-$1200~km~s$^{-1}$ and its velocity dispersion ranges from 400-800~km~s$^{-1}$. In general, this is the same as that of the Ne emission line reported in \citetalias{10.1093/mnras/stac1289}.

\subsubsection{\ion{Fe}{XVII}, \ion{Fe}{XVIII}, and \ion{Fe}{XX} L lines}
Spectral fitting is performed by adopting bvvapec*tbabs using the data for the energy bands of \ion{Fe}{xvii,xviii,xx} emission lines. We set the temperature, Fe abundance, line-of-sight velocity and its dispersion, emission measure of the bvvapec model, and hydrogen column density of the tbabs model as free parameters  and fix the abundances of the other elements at the values obtained with the full energy-band fit \citepalias[Table 4 in][]{10.1093/mnras/stac1289}. The energy bands used are 0.70-0.77~keV (\ion{Fe}{xvii}), 0.81-0.90~keV (\ion{Fe}{xviii}), and 0.93-1.00~keV (\ion{Fe}{XX}).
The results are summarised in Table~\ref{tab:Feallbestfit} and Fig.~\ref{fig:fitOFe} (right). 
The line-of-sight velocity ranges $-$from 800 to 1400~km~s$^{-1}$ and its dispersion ranges
from 500 to 1100~km~s$^{-1}$. These are the same as those measured with the O lines (Section 3.1.1) and Ne lines \citepalias{10.1093/mnras/stac1289}.
\begin{table*}
	\centering
	\caption{Best-fit parameters of the lines of \ion{Fe}{XVII, XVIII, XX} with the bvvapec and tbabs models [see Fig.~\ref{fig:fitOFe}(right) for the plot]. Parameters of abundances other than Fe and $N_{\rm H}$ are fixed at best-fit parameters in the 0.325-5.35 keV band \citepalias[Table 4 in][]{10.1093/mnras/stac1289}. The other parameters are treared the same as in Table~\ref{tab:O78bestfit}. \label{tab:Feallbestfit}}
	\begin{tabular}{lcccccccccccr}
		\hline
		ID & K & A & L & B & D\\
		Phase &0.816&0.912&0.935&0.968&0.987\\
		\hline
		$kT$ (keV) & 0.845$^{+0.048}_{-0.030}$ & $0.844^{+0.051}_{-0.039}$ & 0.834$^{+0.032}_{-0.054}$ & $0.892^{+0.025}_{-0.028}$ & $0.891^{+0.041}_{-0.044}$ \\
		$Z_{\rm O}$ & 1.062 (fixed) & 0.871 (fixed) & 0.848 (fixed) & 0.702 (fixed) & 0.968 (fixed)\\
		$Z_{\rm Ne}$ & 0.775 (fixed) & 0.624 (fixed) & 0.598 (fixed) & 0.586 (fixed) & 1.190 (fixed) \\
		$Z_{\rm Mg}$ & 0.133 (fixed) & 0.117 (fixed) & 0.130 (fixed) & 0.122 (fixed) & 0.316 (fixed)\\
		$Z_{\rm Si}$ & 0.131 (fixed) & 0.140 (fixed) & 0.146 (fixed) & 0.140 (fixed) & 0.664 (fixed)\\
		$Z_{\rm Fe}$=$Z_{\rm Ni}$ & 0.060$^{+0.022}_{-0.014}$ & 0.054$^{+0.025}_{-0.012}$ & 0.060$^{+0.017}_{-0.019}$ & 0.087$^{+0.025}_{-0.020}$ & 0.102$^{+0.037}_{-0.028}$\\
		$v_{\rm los}$ (km~s$^{-1}$) & $-1366^{+236}_{-327}$ & $-962^{+225}_{-199}$ & $-1478^{+162}_{-143}$ & $-1381^{+153}_{-142}$ & $-796^{+183}_{-165}$ \\
		$\sigma_{\rm obs}$ (km~s$^{-1}$) & 1052$^{+482}_{-232}$ & $687^{+295}_{-190}$ & 541$^{+208}_{-183}$ & $630^{+195}_{-161}$ & $570^{+209}_{-166}$ \\
		Norm. ($10^{-14}{\rm EM}/4\pi D^2$ cm$^{-5}$) & 0.027$^{+0.008}_{-0.008}$ & $0.039^{+0.013}_{-0.013}$ & 0.049$^{+0.029}_{-0.013}$ & $0.053^{+0.020}_{-0.013}$ & $0.016^{+0.008}_{-0.005}$ \\
		$N_{\rm H}$ ($10^{22}$ cm$^{-2}$) & 0.284$^{+0.059}_{-0.078}$ & 0.263$^{+0.064}_{-0.086}$  & 0.330$^{+0.103}_{-0.064}$ & 0.333$^{+0.067}_{-0.056}$ & 0.276$^{+0.008}_{-0.005}$  \\
						\hline
		C-statistics (dof) & 2455.89 (2194) & 2475.64 (2149) & 2316.98 (2178) & 2427.54 (2176)& 2482.05 (2180) &\\
		\hline
	\end{tabular}
\end{table*}
\begin{figure*}
	\includegraphics[width=0.7\textwidth]{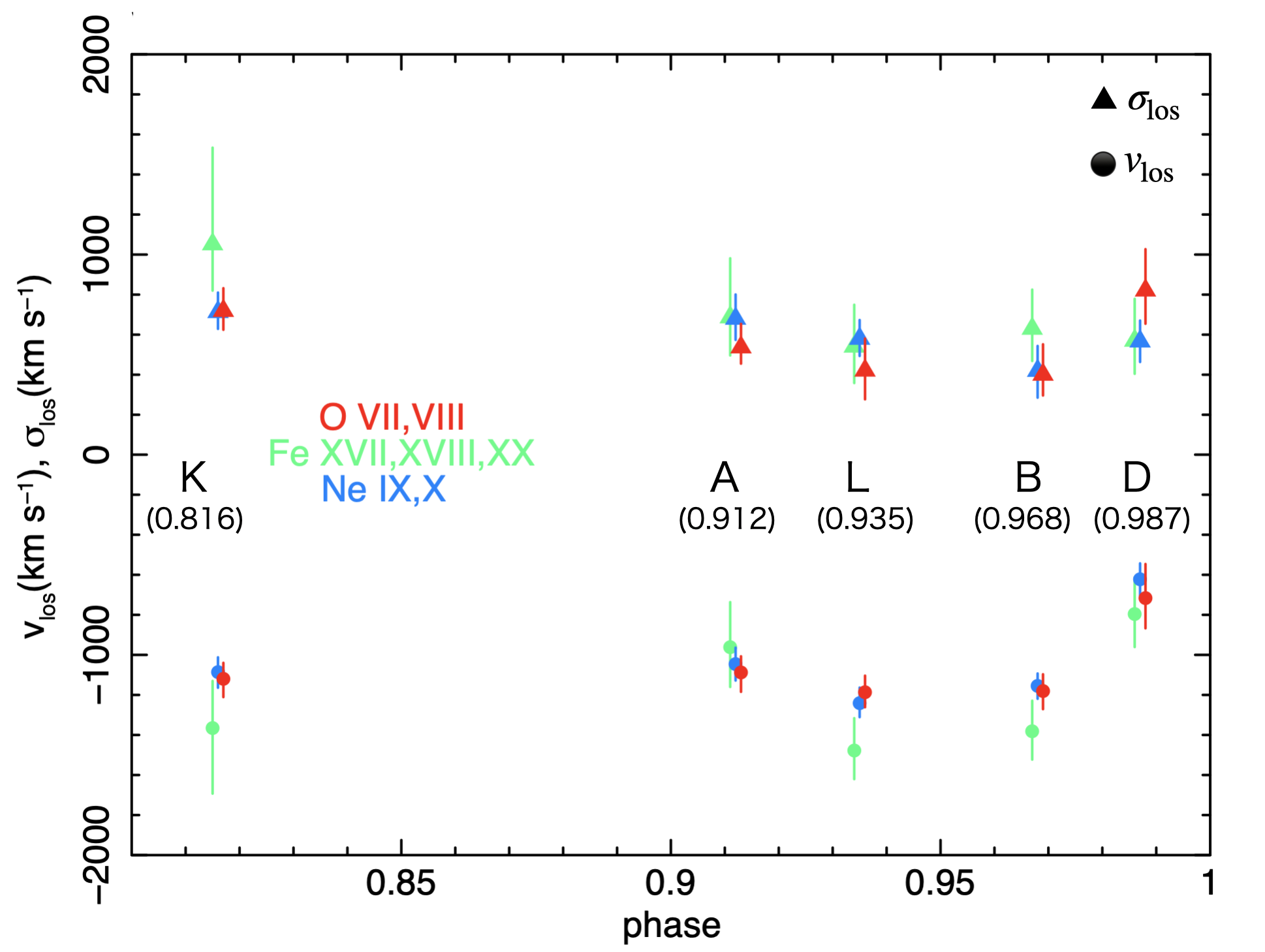}
	\caption{Line-of-sight velocity ($v_{\rm los}$) and its dispersion ($\sigma _{\rm los}$) of \ion{Ne}{ix, x} \citepalias[Table 5 of][]{10.1093/mnras/stac1289}, \ion{O}{vii, viii} and \ion{Fe}{xvii, xviii, xx} (Tables \ref{tab:O78bestfit} and \ref{tab:Feallbestfit}, respectively) plotted as a function of the orbital phase. At all phases, the line-of-sight velocity is negative (blue-shifted), implying that the plasma emitting these lines is approaching the earth. Both $v_{\rm los}$ and $\sigma _{\rm los}$ seem to take their minimum values between phase L (0.935) and B (0.968), where the inferior conjunction of the O star occurs.
	}
	\label{fig:Phase-v}
\end{figure*}

\subsubsection{Summary of the line-of-sight velocity and its dispersion}
We plot the line-of-sight velocity and its dispersion of the Ne, O, and Fe lines as functions of the orbital phase, as displayed in Fig.~\ref{fig:Phase-v}. 
Here, we use the results of Ne from Table 5 reported in \citetalias{10.1093/mnras/stac1289}.
The line-of-sight velocities are blue-shifted at these five phases where the the observer views the collision from within the shock cone.
If these plasmas flow along the shock cone, the line-of-sight speed should be the largest and the velocity dispersion should be the smallest when the observer is closest to the axis of symmetry of the shock cone. The trends shown in Fig. \ref{fig:Phase-v} follow this expectation well, as the inferior conjunction of the O star occurs between phases B (0.968) and L (0.935). 
The O and Fe lines follow a trend similar to that of the Ne lines, reinforcing the results of Ne lines reported in \citetalias{10.1093/mnras/stac1289}. Note that we neglect the effect of the Coriolis force, which is not sufficiently strong to affect the axial symmetry of the shock cone at the phases before periastron passage (see APPENDIX A for the effect of the Coriolis force on the shape of the shock cone). 

\subsection{Density diagnosis}
\subsubsection{He-like triplet of Oxygen}

As explained in \citetalias{10.1093/mnras/stac1289} \S3.4, and \S4.2.1, the intensity ratio of the He-like triplet lines from the heavy elements is sensitive to the plasma density. Following the analysis method of Ne reported in \citetalias{10.1093/mnras/stac1289}, we attempt to constrain the plasma electron number density $n_{\rm e}$ with a He-like triplet of O. 

We use the energy band of the He-like triplet of \ion{O}{vii} (0.55-0.59 keV; see Fig. \ref{fig:fitOFe} left). 
As a continuum, we adopt the model composed of bvvapec multiplied by tbabs. 
We append three velocity-shifted gaussians (zgauss) 
on this continuum to represent $f$, $i$, and $r$ components of \ion{O}{vii}, and, instead, fix the O abundance of the bvvapec model to 0. The other parameters, including $N_{\rm H}$ of the tbabs model, are obtained using the full energy band fit \citepalias[][Table 4]{10.1093/mnras/stac1289}. 
We fixed the centroid energies of the $f$, $i$, and $r$ components at their rest-frame energies (0.5610 keV, 0.5687 keV, and 0.5740 keV, respectively), and their
velocity shift is realized with the common redshift parameter of the zgauss models, which is $v_{\rm los} (= cz)$ shown in Table~\ref{tab:O78bestfit}.
The energy width $\sigma$ [keV] of the forbidden line is linked to $\sigma_{\rm los}$ [km~s$^{-1}$] of bvvapec listed in Table \ref{tab:O78bestfit} through 
$\sigma_{\rm f}$=($\sigma_{\rm los}$/$c$)$E_{\rm f}$, where $E_{\rm f}$ is the forbidden-line central energy.
$\sigma_{\rm i}$ and $\sigma_{\rm r}$ are scaled with $\sigma_{\rm f}$ according to their line central energies.

Even with these constraints, as reported in \citetalias{10.1093/mnras/stac1289} \S3.4, 
we evaluate the uncertainty of the line parameters associated with $v_{\rm los}$ manually.
We adopt the errors of $v_{\rm los}$ determined by the K$\alpha$ lines of \ion{O}{vii} and \ion{O}{viii} (Table~\ref{tab:O78bestfit}).
First, we perform spectral fitting at the best-fit $v_{\rm los}$ value. We then repeat the same fit at the maximum/minimum values of the confidence interval to obtain the errors in the line parameters. 
The same procedure is performed for all five phases. The resulting intensities of the triplet components are summarised in Table \ref{tab:O_tripletfit}.

\begin{table*}
	\centering
	\caption{Best-fit parameters of the He-like triplet of O using the model comprised of bvvapec and three zgauss components, which represent a resonance line, intercombination lines, and a forbidden line. Instead, the O abundance of bvvapec is set to zero, and the other parameters of bvvapec and tbabs are set to the best-fit values obtained from fitting with \ion{O}{vii} and \ion{O}{viii} lines (Table \ref{tab:O78bestfit}). The model parameter redshift $v_{\rm los}$/$c$ is not constrained very well solely with the He-like line. Hence, the fitting is performed with $v_{\rm los}$ fixed at its best-fit value and at the minimum and maximum values of the confidence interval separately. The line centroid energies $E_{\rm f}$, $E_{\rm i}$, and $E_{\rm r}$ are fixed at the rest-frame energies of the He-like triplet (0.561, 0.569, and 0.574 keV, respectively) and effectively
    set free to vary with the common value of the redshift ($v_{\rm los}/c$).  
    $\sigma_{\rm f}$ is fixed at $\sigma_{\rm f} = (\sigma_{\rm obs}/c) E_{\rm f}$, and $\sigma_{\rm i}$ and
    $\sigma_{\rm r}$ are linked to $\sigma_{\rm f}$ with their line energy ratios ($E_{\rm i}$/$E_{\rm f}$ and $E_{\rm r}$/$E_{\rm f}$, respectively) As a result, the free parameters of the fitting is the intensities of the lines.
    \label{tab:O_tripletfit}}
{\tabcolsep = 2pt
\begin{tabular}{lccccccc}
		\hline
                & & $v_{\rm los}$ & $\sigma_{\rm f}$&Norm.$_{\rm f}$ &Norm.$_{\rm i}$&Norm.$_{\rm r}$ & C-statistics\\
            Phase& & (km~s$^{-1}$)& (10$^{-3}$ keV) &  ($10^{-5}$ s$^{-1}$ cm$^{-2}$)& ($10^{-5}$ s$^{-1}$ cm$^{-2}$) & ($10^{-5}$ s$^{-1}$ cm$^{-2}$) & (14 bins)
 \\ \hline \hline 
             & min& $-$1202 (fixed) & &1.984$^{+1.220}_{-1.001}$& 0.946$^{+1.150}_{-0.718}$ & 2.463$^{+1.286}_{-1.095}$ &4.30 \\
            K (0.816)& best-fit& $-$1119 (fixed)&1.131 (fixed)&2.000$^{+1.296}_{-0.951}$& 0.930$^{+1.150}_{-0.733}$&2.485$^{+1.294}_{-1.057}$&3.77\\
             &max & $-$1028 (fixed)& &2.012$^{+1.272}_{-0.967}$ & 0.907$^{+1.248}_{-0.700}$ & 2.503$^{+1.340}_{-1.001}$ &3.37\\ \hline
             & min& $-$1185 (fixed)& &2.234$^{+1.748}_{-1.310}$ &2.051$^{+1.778}_{-1.268}$&4.324$^{+2.291}_{-1.956}$&21.12\\
            A (0.912)&best-fit&$-$1088 (fixed)& 1.004 (fixed)& 2.214$^{+1.815}_{-1.316}$& 1.931$^{+1.884}_{-1.218}$&4.405$^{+2.442}_{-2.054}$&22.55\\
             &max & $-$1007 (fixed)& &2183$^{+1.695}_{-1.396}$& 1.803$^{+1.877}_{-1.308}$ &4.474$^{+2.449}_{-1.983}$ &23.82\\ \hline
             & min&$-$1261 (fixed)& &2.098$^{+1.847}_{-1.435}$&0.262$^{+1.597}_{-0.202}$ &3.466$^{+1.743}_{-1.532}$ &11.44\\
            L (0.935)& best-fit& $-$1186 (fixed)&0.749 (fixed)&2.227$^{+1.802}_{-1.452}$&0.182$^{+1.644}_{-0.120}$& 3.543$^{+1.696}_{-1.660}$&10.22\\
             &max &$-$1103 (fixed)& & 2.346$^{+1.883}_{-1.479}$ &0.097$^{+1.700}_{-0.043}$& 3.614$^{+1.741}_{-1.633}$&9.05\\ \hline
             & min& $-$1271 (fixed)& &1.269$^{+1.226}_{-0.796}$&1.075$^{+1.342}_{-0.784}$&  2.480$^{+1.506}_{-1.210}$&6.81\\
            B (0.968)& best-fit& $-$1180 (fixed)& 0.821 (fixed)&1.257$^{+1.240}_{-0.801}$&1.042$^{+1.318}_{-0.771}$&2.473$^{+1.555}_{-1.241}$&7.39\\
             &max & $-$1096 (fixed)& &  1.238$^{+1.161}_{-0.826}$&1.013$^{+1.251}_{-0.769}$&2.451$^{+1.543}_{-1.229}$&8.11 \\ \hline
             & min& $-$867 (fixed)& &0.199$^{+0.854}_{-0.158}$&$<$ 0.077 &1.299$^{+0.789}_{-0.937}$& 11.08\\
            D (0.987)& best-fit& $-$716 (fixed)& 1.536 (fixed)& 0.205$^{+0.789}_{-0.172}$&$<$ 0.826&1.357$^{+0.683}_{-1.024}$&11.62\\
             &max & $-$546 (fixed)& & 0.168$^{+0.792}_{-0.135}$&$<$ 0.873&1.288$^{+0.720}_{-0.981}$&12.34\\     
		\hline \end{tabular}
		}
\end{table*}

\begin{figure}
    \includegraphics[width=.45\textwidth]{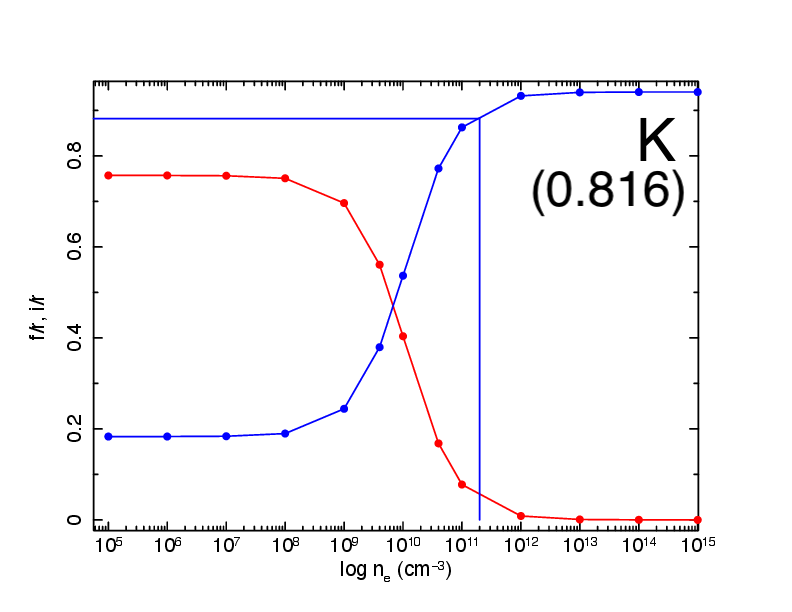}
    \includegraphics[width=.45\textwidth]{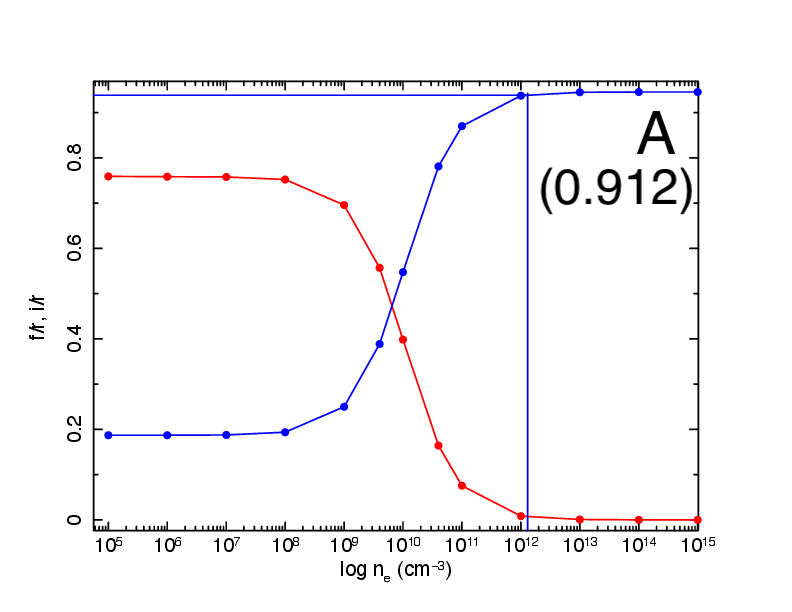}
    \includegraphics[width=.45\textwidth]{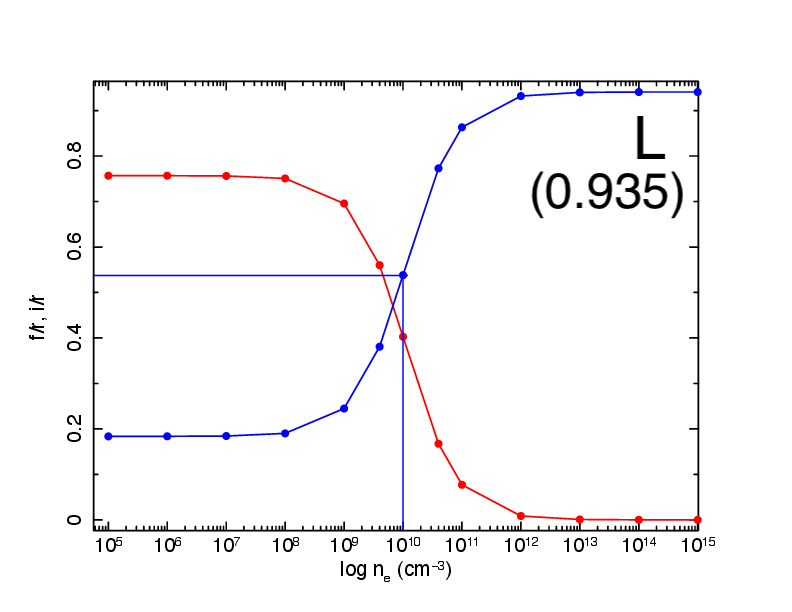}
    \caption{Density diagnosis of the plasma with the intensity ratio of the He-like triplet of \ion{O}{vii} at phases K (0.816), A (0.912), and L (0.935).
    The red and blue curves represent $f/r$ and $i/r$, respectively. 
    The vertical axis is the intensity ratio and the horizontal axis is the electron number density. The blue boxes represent the observed ratio $i/r$ in the vertical axis and the resultant density upper limits in the horizontal axis. Since the allowed range of the observed $f/r$ ratio is wider than the theoretical range shown in the vertical axis, we cannot pose any constraint on the density from the $f/r$ ratio. The electron number density at phases B (0.968) and D (0.987) could not be constrained with this method.     
        \label{fig:O_ne}}
\end{figure}

In parallel to the analysis of the spectra, as reported in \citetalias{10.1093/mnras/stac1289} \S4.2.1, we calculate the intensities of $f$, $i$, and $r$ of \ion{O}{vii} using the plasma code SPEX version 3.06.01\footnote{https://www.sron.nl/astrophysics-spex} developed by SRON\footnote{https://www.sron.nl} \citep{1996uxsa.conf..411K} as a function of the electron number density $n_{\rm e}$ and plot the ratio of $f/r$ and $i/r$. We assume that the plasma is in collisional ionisation equilibrium and use the CIE model.
We use the resonance line $r$ to help constrain the density if either $f$ or $i$ is weak.
By comparing the SPEX curves with the intensity ratios derived from the fitting in Table \ref{tab:O_tripletfit}, we are potentially able to determine the electron number density $n_{\rm e}$. The results obtained under the assumption of pure collisional excitation are shown in Fig.~\ref{fig:O_ne}.
Based on this figure, the upper limits of $n_{\rm e}$ are obtained at phases K (0.816), A (0.912), and L (0.935), which are summarised in Table~\ref{tab:O_ne}.
However, these upper limits may be conservative depending on the degree of photo-ionisation effect due to the EUV light from the O star, which is evaluated in the next subsection.
\begin{table}
	\centering
	\caption{Electron number density $n_{\rm e}$ obtained with the He-like triplet of O. We obtain only the upper limits at phases K (0.816), A (0.912), and L (0.935) (Fig. \ref{fig:O_ne}) and no constraint on the other two phases. Note that the EUV excitation overwhelms the collisional excitation effect at phase L (0.935) (Table \ref{tab:UVrate}); hence, its upper limit is extremely conservative. 
 \label{tab:O_ne}}
	\begin{tabular}{ccc} 
		\hline
		Phase & $n_{\rm e}$ (cm$^{-3}$)  \\
		\hline
		K (0.816)& $<$2.0 $\times$ 10$^{11}$  \\
		A (0.912)& $<$1.3 $\times$ 10$^{12}$ \\
		L (0.935)& $<$1.0 $\times$ 10$^{10}$ \\
		B (0.968)& - \\
		D (0.987)& - \\
				\hline
	\end{tabular}
\end{table}

\subsubsection{EUV radiation effect on the O and Ne line-emission site densities}
Using the electron number densities $n_{\rm e}$ obtained in the previous subsection (Table~\ref{tab:O_ne}), 
we calculate the rates of collisional excitation $n_{\rm e}q_{ij}$ and photoexcitation $\Gamma_{ij}$ from the $^3$S level to $^3$P of oxygen
using eqs. (19) and (20) in \citetalias{10.1093/mnras/stac1289}, respectively, together with the ratios of the latter to the former: 
The results are summarised in Table~\ref{tab:UVrate}.
\begin{table}
	\centering
	\caption{
    Rates of collisional excitation and photo-excitation from the $^3$S level to $^3$P, and the ratio of the latter to the former. For Ne, we recalculate the values using the updated line-emission site locations, to be presented in Section 4.1.2.
    \label{tab:UVrate}}
    \setlength{\tabcolsep}{4pt}
	\begin{tabular}{ccccc} 
		\hline
		& Phase & $n_{\rm e}q_{ij}(T_{\rm e})$ (s$^{-1}$)& $\Gamma_{ij} (T_{\rm r},\nu_{ij})$ (s$^{-1}$)& fraction $({\rm per cent})$ \\
		\hline
    & K (0.816)& ~~~~~$<$ 2330 & 70 - 210 & ~~~$>$ 2.9 \\
		& A (0.912)&  ~~~~~~~$<$ 15900 & 210 - 630 & ~~~$>$ 1.3 \\
		O & L (0.935)& ~~~$<$ 120& 270 - 820 & ~~~~~~~$>$ 233.4 \\
		& B (0.968)& - & - & - \\
		& D (0.987)& - & - & -  \\
				\hline
		& K (0.816)& 4580 - 27600 & 170 - 520 & 0.6 - 11.5 \\
		& A (0.912)& 2950 - 16800 & 540 - 1540 & 3.2 - 52.3 \\
		Ne & L (0.935)& ~~~~~~~$<$ 26800& 560 - 1670 & ~~~$>$ 2.1 \\
		& B (0.968)& ~~~~~$<$ 7230 & 1260 - 3820 & ~~~~~$>$ 17.5 \\
		& D (0.987)& ~~~~~~~$<$ 25800 & 12420 - 32360 & ~~~~~$>$ 48.1  \\
				\hline
	\end{tabular}
\end{table}
Note that we use the distances to the line-emission sites calculated later in \S4.1.2 (Table~\ref{tab:NeOlocation}).
For oxygen, because only the upper limits of $n_{\rm e}$ are determined (Table~\ref{tab:O_ne}), the rates of collisional excitation are the upper limits, and only the lower limits are determined for the UV excitation fractions. 
At phases K (0.816) and A (0.912), we have lower limits of a few per cents. In contrast, at phase L (0.935), the high fraction of EUV excitation implies that the enhancement of the intercombination lines $i$ is realised with the EUV radiation from the O star. Hence, the upper limit of $n_{\rm e}$ is very conservative compared to those at phases K (0.816) and A (0.912).

The density upper limit of the O-line emission site is consistent with that of Ne at phase A (0.912), whereas
this is smaller than the allowable range of the Ne line emission site density \citepalias[0.47-2.83$\times$10$^{12}$~cm$^{-3}$,][]{10.1093/mnras/stac1289} at phase K (0.816).
However, the emission sites of these two elements should be different in position along the shock cone. The emissivity of the He-like K$\alpha$ line of O peaks at a temperature of 0.18~keV ($\simeq$2 MK), whereas that of the Ne peaks at $kT = 0.33$~keV \citep[$\simeq$4~MK, ][]{1985A&AS...62..197M}. 
It has been debated whether clumps in the plasma develop or deteriorate along the shock cone. In fact, \citet{1992ApJ...386..265S} pointed out that plasma clumps can develop spontaneously after the plasma experiences shock owing to plasma instability. 
In contrast, dense clumps that may exist in the pre-shock stellar wind may be rapidly destroyed after entering the collision shock \citep{2007ApJ...660L.141P}. We require a much higher quality spectrum of oxygen K$\alpha$ emission lines to form a clear conclusion on the density profile along the shock cone.

Using the updated distance to the Ne line-emission sites to be presented in Table~\ref{tab:NeOlocation}, the effect of EUV radiation on the Ne line-emission site densities is updated and also summarized in Table~\ref{tab:UVrate}.
Because the line-emission sites are more distant than those reported in \citetalias{10.1093/mnras/stac1289}, the 
photoexcitation probabilities are lower than those shown in Table~10 of \citetalias{10.1093/mnras/stac1289}. 
This effect appears most remarkably at phase K (0.816), where the contribution of the photoionisation effect becomes $\lesssim$10 per cent of the collisional excitation (Table~\ref{tab:UVrate}). We believe that the Ne line-emission site density 0.47-2.83$\times$10$^{12}$~cm$^{-3}$ \citepalias{10.1093/mnras/stac1289} is more plausible.

\subsubsection{\ion{Fe}{xvii} L lines}
The intensities of some Fe-L lines are sensitive to the electron number density of the plasma. 
For \ion{Fe}{xvii} (Ne-like Fe), the density-sensitive lines are 0.7242 keV (17.10 \AA) and 0.7263 keV (17.05 \AA) \citep{2001ApJ...560..992M}, with an intensity of 17.05 {\AA} stronger and that of 17.10 {\AA} weaker as the electron number density increases. 
In addition to these two lines, a density-insensitive line at 0.7382 keV (16.78 \AA) is also observed, as shown in Fig. \ref{fig:fitOFe}, although the energy resolution of the RGS is not high enough to fully resolve them. 

We attempt to evaluate $n_{\rm e}$ using the ratios $I$(17.05 \AA)/$I$(16.78 \AA) and $I$(17.10 \AA)/$I$(16.78 \AA) \citep{2001ApJ...560..992M}.
Spectral fitting is performed using the energy bands of the \ion{Fe}{xvii} lines (0.70-0.77 keV). 
In contrast to the case of O (Section 3.2.1), as Fe contributes not only to the line emission but also continuum emission, we are not able to set the Fe abundance equal to 0. Instead, we remove these three emission lines from the bvvapec model and added three 
velocity-shifted Gaussian (zgauss) to the fitting model. 
As in the analysis of \ion{O}{vii}, the line central energies are set free to vary using the redshift parameter $v_{\rm los}$ that is common among the three lines. The line widths are constrained to vary in proportion to the line central energies.

We calculate the ratios $I$(17.05 \AA)/$I$(16.78 \AA) and $I$(17.10 \AA)/$I$(16.78 \AA) with the intensities derived and compare them with the theoretical curves as a function of the electron number density $n_{\rm e}$ \citep{2001ApJ...560..992M}.
However, the intensity ratios are not constrained 
at all; the theoretical ratios $I$(17.05 \AA)/$I$(16.78 \AA) and $I$(17.10 \AA)/$I$(16.78 \AA) vary in the range from 1.1 to 1.5 and from 0.0 to 0.9, respectively, as a function of $n_{\rm e}$ \citep[][Figure 2]{2001ApJ...560..992M}, whereas
the observed ratios at phase L (0.935), which have the best statistics among the five data sets, ranged from 0.89 to 2.79 and from 0.00 to 1.21, respectively. 
This is because the line separations are comparable to the energy resolution of the RGS. Consequently, we are unable to make any meaningful constraints on $n_{\rm e}$ with the \ion{Fe}{xvii} lines.

\section{Discussion}

\subsection{Line-of-sight velocity and its dispersion along the shock cone}

In \citetalias{10.1093/mnras/stac1289}, we argued that the ratio of the observed line-of-sight velocity and its dispersion ($|v_{\rm los}|/\sigma _{\rm los}$) provides a reliable estimate of the line-emission site as long as the plasma flow is laminar, because in this case, $|v_{\rm los}|/\sigma _{\rm los}$ is a monotonically increasing function of the distance from the stagnation point and its value is uniquely determined by the shock cone geometry, which is calculated based on the ram pressure balance between the stellar winds from the two stars. 
However, if the plasma flow includes a turbulent component, the observed $\sigma _{\rm los}$ (= $\sigma_{\rm obs}$) would be enhanced, as expressed in Equation~(23) in \citetalias{10.1093/mnras/stac1289}. 
In addition, if the line-emission sites extend spatially along the shock cone, the variation of $v_{\rm los}$ within the sites further enhances $\sigma_{\rm obs}$ (this possibility is not considered in \citetalias{10.1093/mnras/stac1289}).
Thus, the ratio $|v_{\rm los}|/\sigma _{\rm los}$ tends to underestimate the distance of the line-emission site from the stagnation point.
However, the observed $v_{\rm los}$, that is, the net plasma velocity on a macro scale, is not affected by the turbulence by definition, and we believe that $v_{\rm los}$ is a more reliable tool for deriving the line-emission sites than $|v_{\rm los}|/\sigma _{\rm los}$. 
Accordingly, in this Section 4.1, we first calculate $v_{\rm los}$ and $\sigma _{\rm los}$ separately along the shock cone.

\subsubsection{Initial velocity at the stagnation point}
To calculate $v_{\rm los}$ and $\sigma_{\rm los}$ along the shock cone, we must calculate the initial velocity of the plasma flow at the stagnation point.
At the stagnation point, the macroscopic velocity is zero because the stellar winds makes head-on collision. In such a case, the plasma is expected to flow out
at the speed of sound. Thus, we calculate the speed of sound based on the fact that the plasma temperature is 3.5 keV (41 MK) at the stagnation point \citep{2015PASJ...67..121S}.

Under the assumption that the abundance of WR stars is He:C = 5:2 \citep{1999ApJ...519..354H} and those of the O star are H:He = 10:1, the mean molecular weights $\mu_{\rm wr}$ and $\mu_{\rm o}$ of each stellar wind, including the electrons, are calculated to be $\mu_{\rm wr} = 1.52$ and $\mu_{\rm o} = 0.61$.
Here we assume that the stellar winds from the WR and O stars become admixed after passing through the shock surface.
First, we derive the particle number density $n_{\rm rw}$ and $n_{\rm o}$ of each stellar wind at the stagnation point from $\dot M = 4\pi r^2 \rho (r) v(r)$ 
and the stellar wind velocity formula

\begin{equation}
v(r)=v_\infty\left(1-\frac{R}{r}\right)^\beta,
\label{eq:v}
\end{equation}
where we set $\beta$=1 not only for the O star but also for WR star. This is acceptable in our case, because the shock cone is formed far away from the WR star, and hence the WR wind velocity there is nearly insensitive to the choice of $\beta$ \citep{2015PASJ...67..121S}.
\citet{1992ApJ...389..635U} stated that the acceleration of winds is almost negligible beyond 3-5 times the radius of the star, and according to \citet{2015PASJ...67..121S}, the braking of the stellar wind of the WR star owing to the EUV radiation from the O star can be ignored because the temperature of the hot component does not decrease until phase D (0.987). 
Table~\ref{tab:par2} summarises the distances of the stagnation point from the two stars and the stellar wind parameters at each phase. 
\begin{table*}
  \centering
  \caption{{Stagnation point distance from each star and the densities and velocities of the pre-shock stellar winds from the two stars at the stagntion point.
  Notably the wind velocity of the WR star reaches its terminal velocity 2860 km~s$^{-1}$ at the stagnation point. \label{tab:par2}}
  }
   {\tabcolsep = 8pt
  \begin{tabular}{lcccccccc} \hline
  Phase & \multicolumn{2}{c}{distance} && \multicolumn{2}{c}{velocity} && \multicolumn{2}{c}{density} \\
 &\multicolumn{2}{c}{(10$^{13}$ cm)} && \multicolumn{2}{c}{(10$^3$ km~s$^{-1}$)} && \multicolumn{2}{c}{(10$^{6}$ cm$^{-3}$)} \\ \cline{2-3} \cline{5-6} \cline{8-9}
    &$r_{\rm wr}$ & $r_{\rm o}$&& $v_{\rm wr}$& $v_{\rm o}$&&$n_{\rm wr}$&$n_{\rm o}$ \\ \hline
    K (0.816) & 25.64  & 5.38 && 2.86 & 3.00 && ~~2.31 & ~~~5.10 \\
    A (0.912) & 16.74  & 3.52 && 2.86 & 2.94 && ~~5.43 & ~~12.20 \\
    L (0.935) & 13.77  & 2.89 && 2.86 & 2.91 && ~~8.02 & ~~18.24 \\
    B (0.968) & ~~8.47 & 1.78 && 2.86 & 2.78 && 21.20  & ~~50.35 \\
    D (0.987) & ~~4.46 & 0.94 && 2.86 & 2.50 && 76.33  & 201.81  \\
    \hline
  \end{tabular}
  }
\end{table*}
The mean molecular weight $\mu$ of the plasma in the shock cone is calculated using the following equation:
\begin{equation}
\mu = \frac{\mu_{\rm wr}n_{\rm wr}v_{\rm wr} +\mu_{\rm o}n_{\rm o}v_{\rm o}}{n_{\rm wr}v_{\rm wr} + n_{\rm o}v_{\rm o}},
\label{eq;mutotal}
\end{equation}
which results in $\mu = $0.88, independent of the orbital phase.
As $\gamma$=5/3, $k_{\rm B}T$=3.5 keV \citep[= 41~MK, ][]{2015PASJ...67..121S}, and $m_{\rm H}$ = 1.67 $\times 10^{-24}$~g, the speed of sound $c_{\rm s}$ is: 
\begin{equation}
  c_{\rm s}=\sqrt{\gamma \frac{k_{\rm B}T}{\mu m_{\rm H}}}=8.0\times10^2 {\rm [km~s^{-1}]},
\label{eq;Eflow}
\end{equation} 
which is adopted as the plasma outflow velocity at the stagnation point.

Thus far, we have assumed that the winds from the WR and O stars become admixed immediately after they experience the shock. 
However, some previous studies claim 
that it takes some time for the winds to mix \citep[e.g.][]{1992ApJ...389..635U,1992ApJ...386..265S}. 
In this case, the two winds have different sound velocities at the stagnation point. According to Equation~(\ref{eq;Eflow}), with $\mu = \mu_{\rm wr}$ and $\mu_{\rm o}$, they are 610 and 960 km~s$^{-1}$, respectively, which differ from the values obtained using Equation~(\ref{eq;Eflow}) only by 20-30 per cent. 
In Section 4.4, we show that this difference barely affects characterisation of the nature of the plasma flow. 

\subsubsection{Line-of-sight velocity and its dispersion}
\begin{figure}
	\includegraphics[width=\columnwidth]{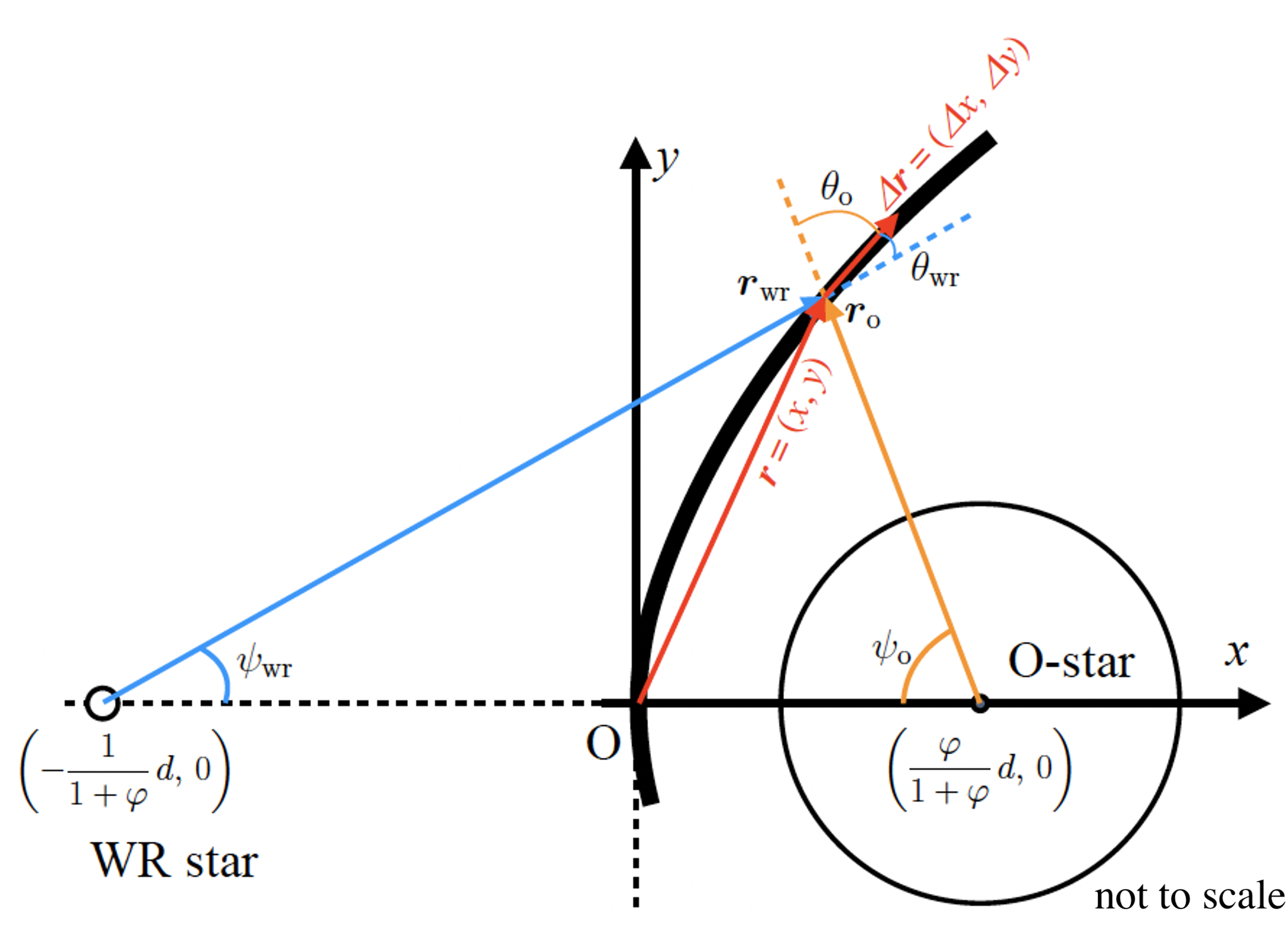}
    \caption{Scheme of calculating the plasma flow velocity along the shock cone. The $x$-axis is the line connecting the centres of the WR and O stars, and the stagnation point is at the origin. 
    Note that $\varphi = \sqrt{ \frac{\dot{M}_{\rm o}v_{\infty, {\rm o}}}{\dot{M}_{\rm wr}v_{\infty, {\rm wr}}} }$. $\theta_{\rm{wr}}$ and $\theta_{\rm o}$ are the angles between the line from each star and the local tangent line of the shock cone. The angles $\psi _{\rm{wr}}$ and $\psi_{\rm o}$ are the angles of those lines measured from the $x$-axis.
    In order to show some parameters clearly, the scaling of the O star is incorrect. See Fig.~10 of \citetalias{10.1093/mnras/stac1289} for the real scaling.}
    \label{fig:zahyou}
\end{figure}

Using the initial velocity of the plasma outflow described in the previous section, we calculate the plasma flow velocity $V$ along the shock cone and transform it into $|v_{\rm los}|$ and $\sigma_{\rm los}$ as functions of the $x$ coordinate in Fig.~\ref{fig:zahyou}. 
We then compare the calculated 
and observed $v_{\rm los}$ to update the locations of the line emission sites.
To derive $v_{\rm los}$ and $\sigma_{\rm los}$ from the flow velocity $V$, we utilise the shape of the shock cone that is determined based on the ram-pressure balance of stellar winds in \citetalias{10.1093/mnras/stac1289} \citep{1992ApJ...389..635U, 1996ApJ...469..729C, 1997MNRAS.292..298P}. 
We consider the coordinate system shown in Fig. \ref{fig:zahyou}.
In this flow configuration, the momentum of the plasma at an arbitrary point \mbox{\boldmath $r$} on the shock cone increases by receiving the tangential component of the momentum of the stellar winds into a small vector $\Delta\bm{r}$ along the shock cone. 
The mass increment rate $\dot M(\mbox{\boldmath $r$})$ and plasma velocity $V(\mbox{\boldmath $r$})$ satisfy the following equations:
\begin{equation}
\begin{split}
\dot M&(\bm{r})V(\bm{r})= \dot M(\bm{r}-\Delta\bm{r})V(\bm{r}-\Delta\bm{r})\\
&+\dot M_{\rm{wr}}\frac{\Delta\Omega_{\rm{wr}}(\bm{r}_{\rm{wr}})}{4\pi}v_{\rm{wr}}(\bm{r}_{\rm{wr}})\cos\theta_{\rm{wr}}
 + \dot M_{\rm o}\frac{\Delta\Omega_{\rm o}(\bm{r}_{\rm o})}{4\pi}v_{\rm o}(\bm{r}_{\rm o})\cos\theta _{\rm o},
	\label{eq:MV}
\end{split}
\end{equation}
where $v_{\rm wr,o}$ and $\theta_{\rm wr,o}$ are the stellar wind velocity and angle between the stellar wind vector and vector $\Delta\bm{r}$, respectively; and $\Delta\Omega_{\rm wr,o}$ is the solid angle subtended by the annulus containing the vector $\Delta\bm{r}$ over each star, expressed as
$\Delta\Omega_{\rm wr,o}(\bm{r})=\Omega_{\rm wr,o}(\bm{r})-\Omega_{\rm wr,o}(\bm{r}-\Delta\bm{r})$,
where 
$\Omega_{\rm wr,o}(\bm{r})= 2\pi(1- \cos\psi_{\rm wr,o})$.
$\dot M_{\rm wr,o}$ is the mass-loss rate of each star. 
The effect of gravity can be ignored here as the escape velocity of the O star's wind at the stagnation point is less than 1/10 of the terminal velocity of the O star's wind $v_{\infty,{\rm o}}$. 
$\dot M(\bm{r})$ is expressed as follows:
\begin{equation}
\begin{split}
\dot M&(\bm{r})= \dot M_{\rm{wr}}\frac{\Omega_{\rm{wr}}(\bm{r}_{\rm{wr}})}{4\pi}
 + \dot M_{\rm o}\frac{\Omega_{\rm o}(\bm{r}_{\rm o})}{4\pi}.
	\label{eq:M_dot}
\end{split}
\end{equation}

Dividing Equation~(\ref{eq:MV}) with Equation~(\ref{eq:M_dot})
yields the following recurrence equation for $V(\bm{r})$:
\begin{equation}
\begin{split}
V(\bm{r})= & \frac{\dot M(\bm{r}-\bm{\Delta r})}{\dot M(\bm{r})}V(\bm{r}-\bm{\Delta r})+\frac{\dot M_{\rm{wr}}}{\dot M(\bm{r})}\frac{\Delta\Omega_{\rm{wr}}(\bm{r}_{\rm{wr}})}{4\pi}v_{\rm{wr}}(r_{\rm{wr}})\cos\theta _{\rm{wr}}\\
&+\frac{\dot M_{\rm o}}{\dot M(\bm{r})}\frac{\Delta\Omega_{\rm o}(\bm{r}_{\rm o})}{4\pi}v_{\rm o}(\bm{r}_{\rm o})\cos\theta _{\rm o}.
\label{eq:V}
\end{split}
\end{equation}
By using the initial value of $V$ (Section 4.1.1), we obtain the velocity $V$ along the shock cone sequentially using Equation~(\ref{eq:V}).

Next, we calculate the line-of-sight velocity and its dispersion from the plasma flow velocity using Equations~(15) and (17) in \citetalias{10.1093/mnras/stac1289}, respectively,
by incorporating $V({\bm r})$ as,
\begin{equation}
v_{\rm los}(x, y, \theta, i, \omega)=\frac{V(\bm{r})\Delta x}{\sqrt{\Delta x^2 +\Delta y^2}}\sin i \sin (\theta+\omega),
\label{eq:vlos}
\end{equation}
\begin{equation}
\sigma _{\rm los}(x, y, \theta, i, \omega)=\frac{V(\bm{r})\Delta y}{\sqrt{\Delta x^2 +\Delta y^2}}\sqrt{\frac{\sin ^2i \cos ^2(\theta+\omega)+\cos ^2i}{2}}.
\label{eq:sigmalos}
\end{equation}
Note that, for the angular averages, the shielding of part of the shock cone by the O star can be neglected because the solid angle of the O star is considerably smaller than the shock cone \citepalias[Fig. 10 of][]{10.1093/mnras/stac1289}. 

Thus far, we have assumed that the plasma flow is laminar in our calculations of $|v_{\rm los}|$ and $\sigma_{\rm los}$. Although the laminar flow does not have any velocity dispersion, the observed emission lines emanate from an annular region on the shock cone whose different parts are seen under different angles.
As a result, we observe different $v_{\rm los}$ values from different portions of the line-emission plasma, which provides non-zero values to the line-of-sight velocity dispersion $\sigma_{\rm los}$, even in the case of laminar flow. The calculated $\sigma_{\rm los}$ value is purely geometrical in origin. This can be understood from the factor $\Delta y / \sqrt{\Delta x^2 + \Delta y^2}$ in Equation~(\ref{eq:sigmalos}) \citepalias[see Section 4.1.4 of][]{10.1093/mnras/stac1289}.

\subsection{Excess of the velocity dispersion and locations of the Ne and O line-emission sites}

\begin{figure*}
  \begin{minipage}{0.33\textwidth}
    \includegraphics[width=\textwidth]{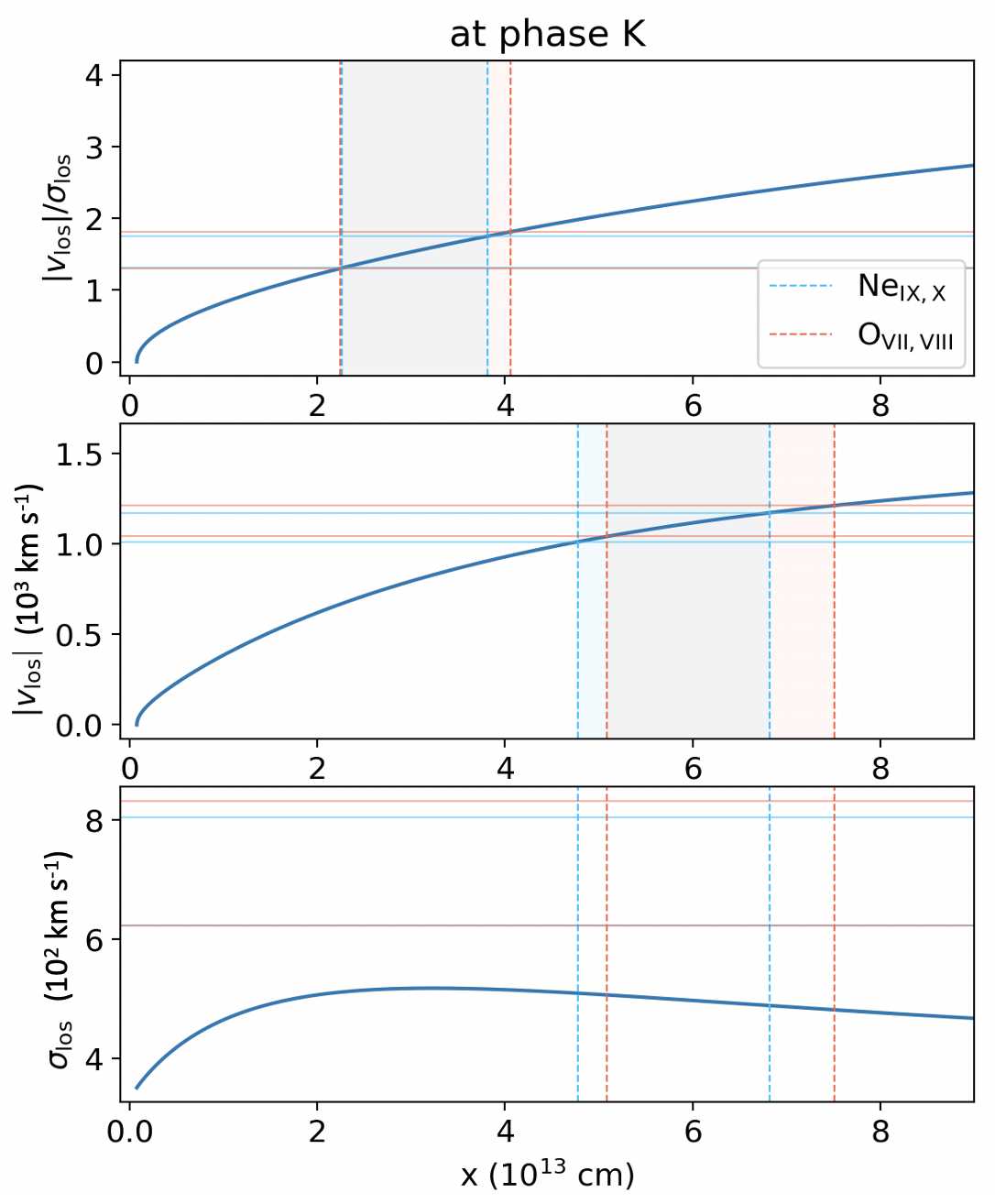}
  \end{minipage}
  \hfill
    \begin{minipage}{0.33\textwidth}
    \includegraphics[width=\textwidth]{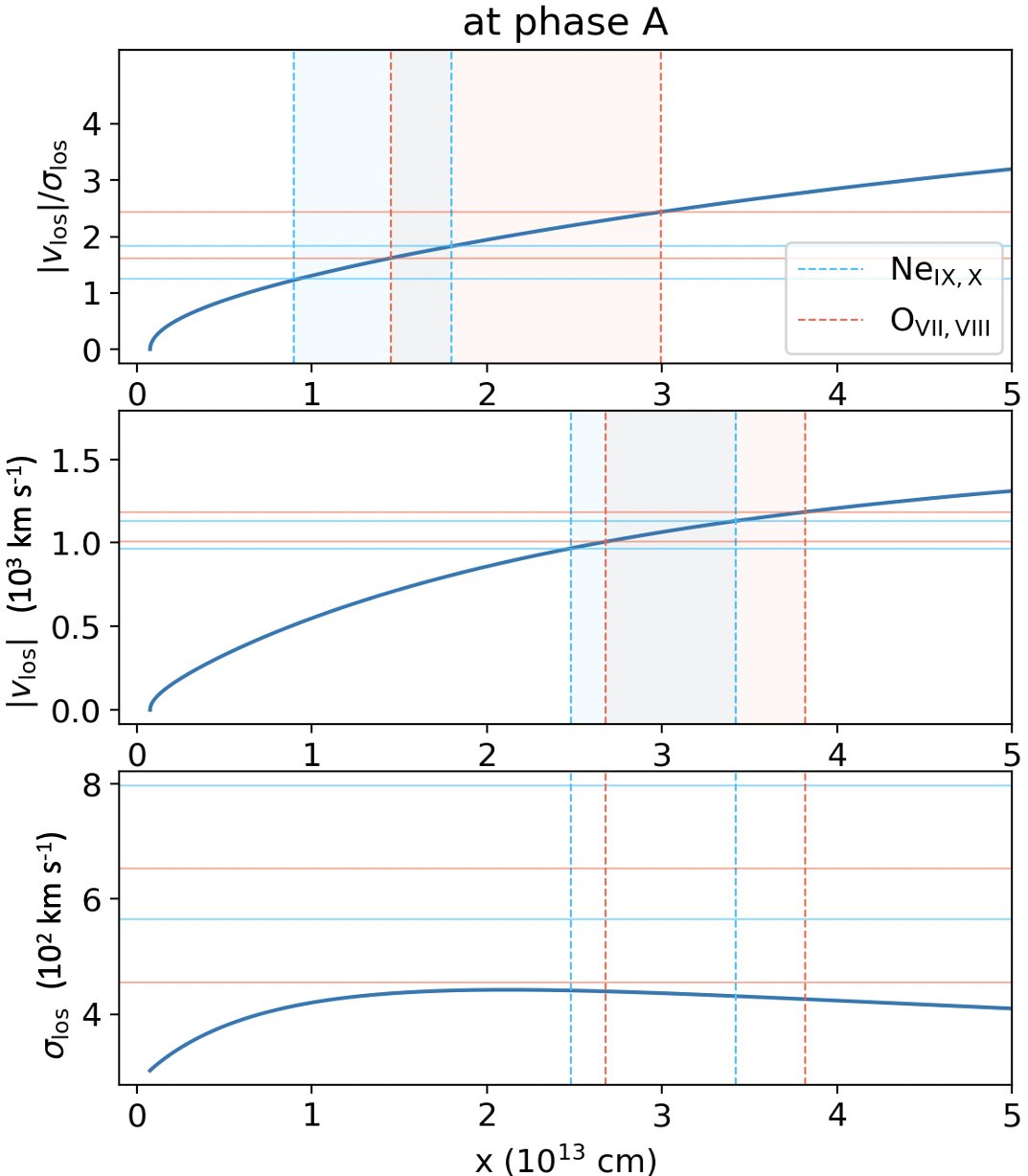}
  \end{minipage}
  \hfill
  \begin{minipage}{0.33\textwidth}
    \includegraphics[width=\textwidth]{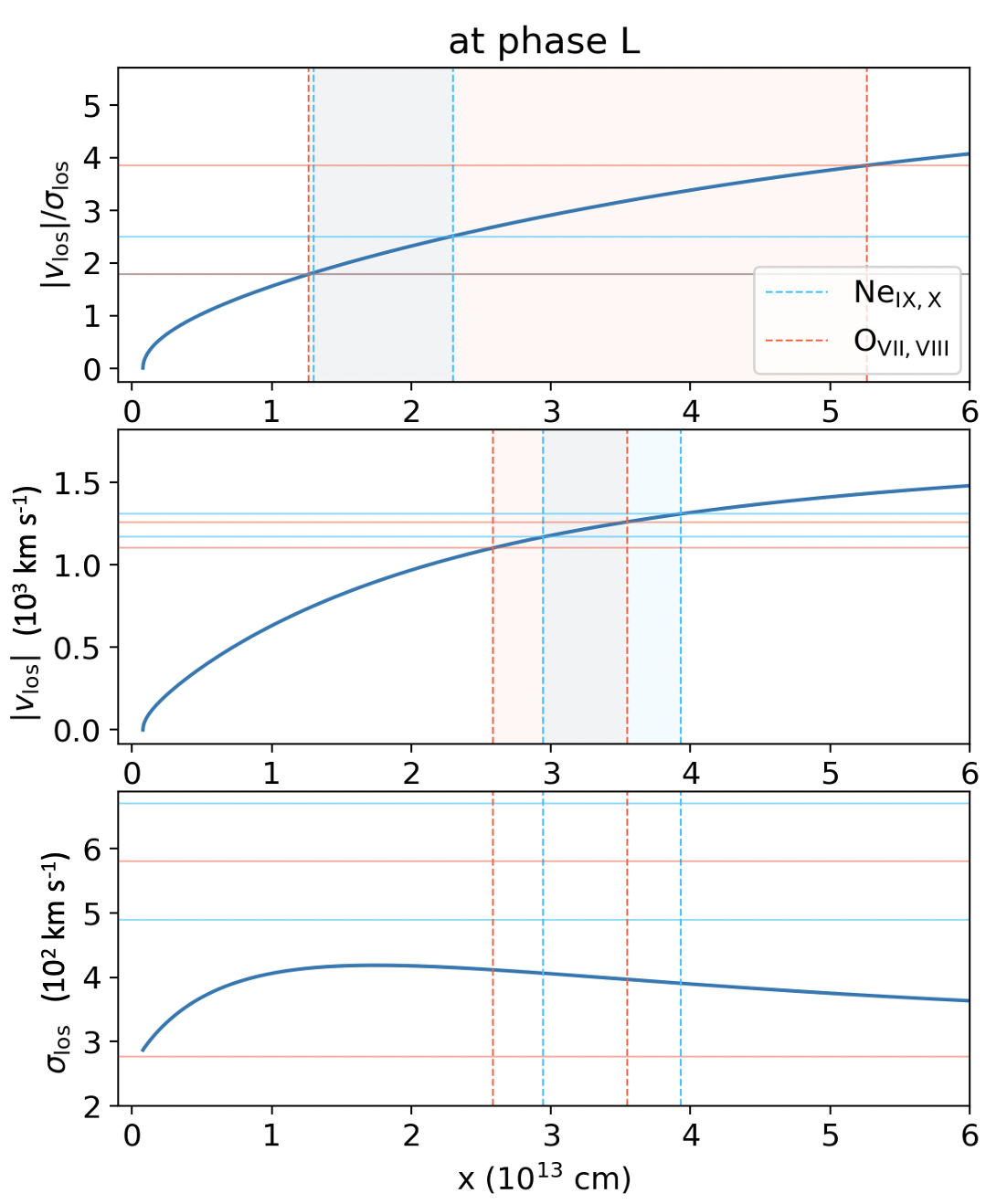}
  \end{minipage}
  \hfill
  \begin{minipage}{0.33\textwidth}
    \includegraphics[width=\textwidth]{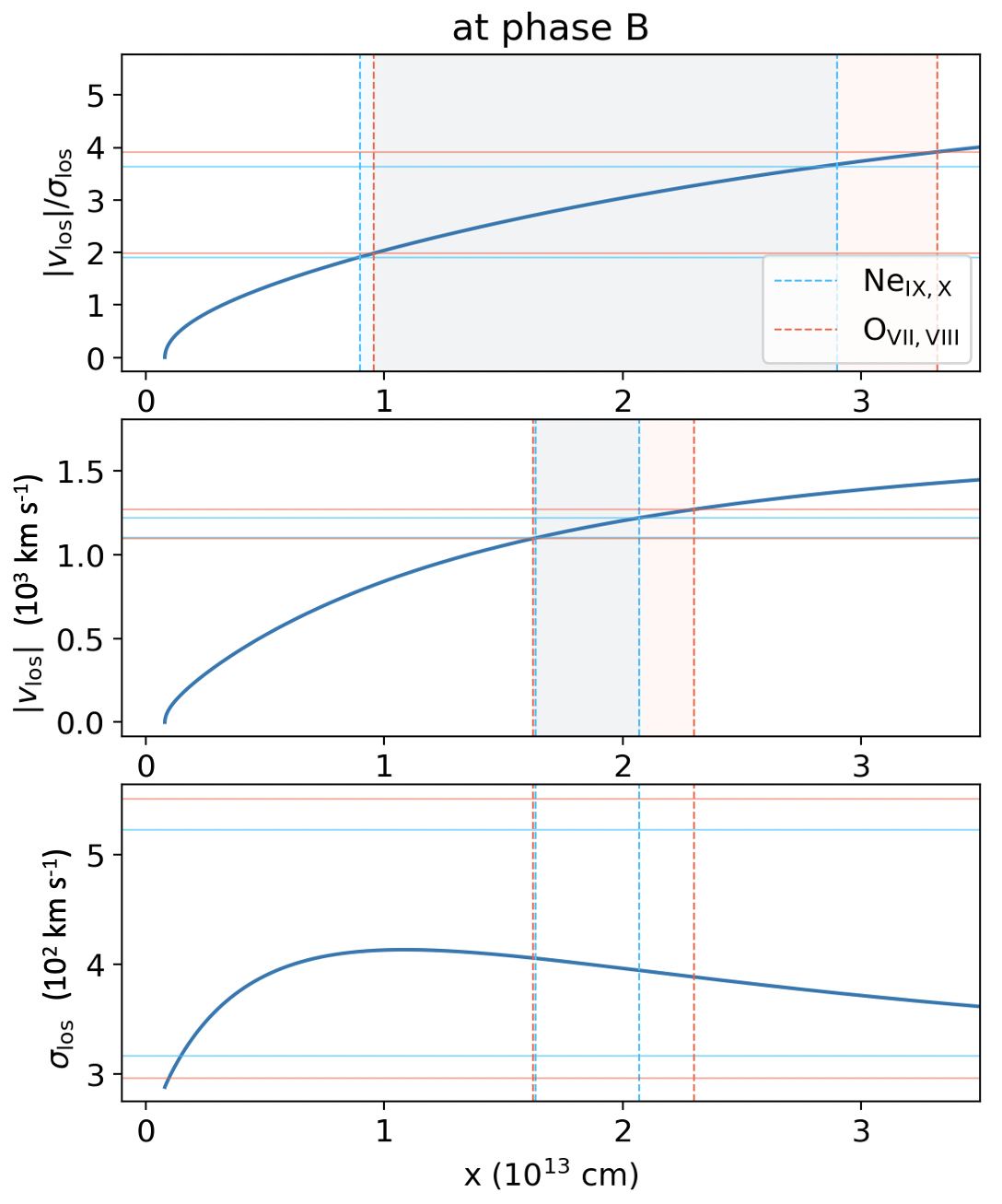}
  \end{minipage}
  \hfill
  \begin{minipage}{0.33\textwidth}
    \includegraphics[width=\textwidth]{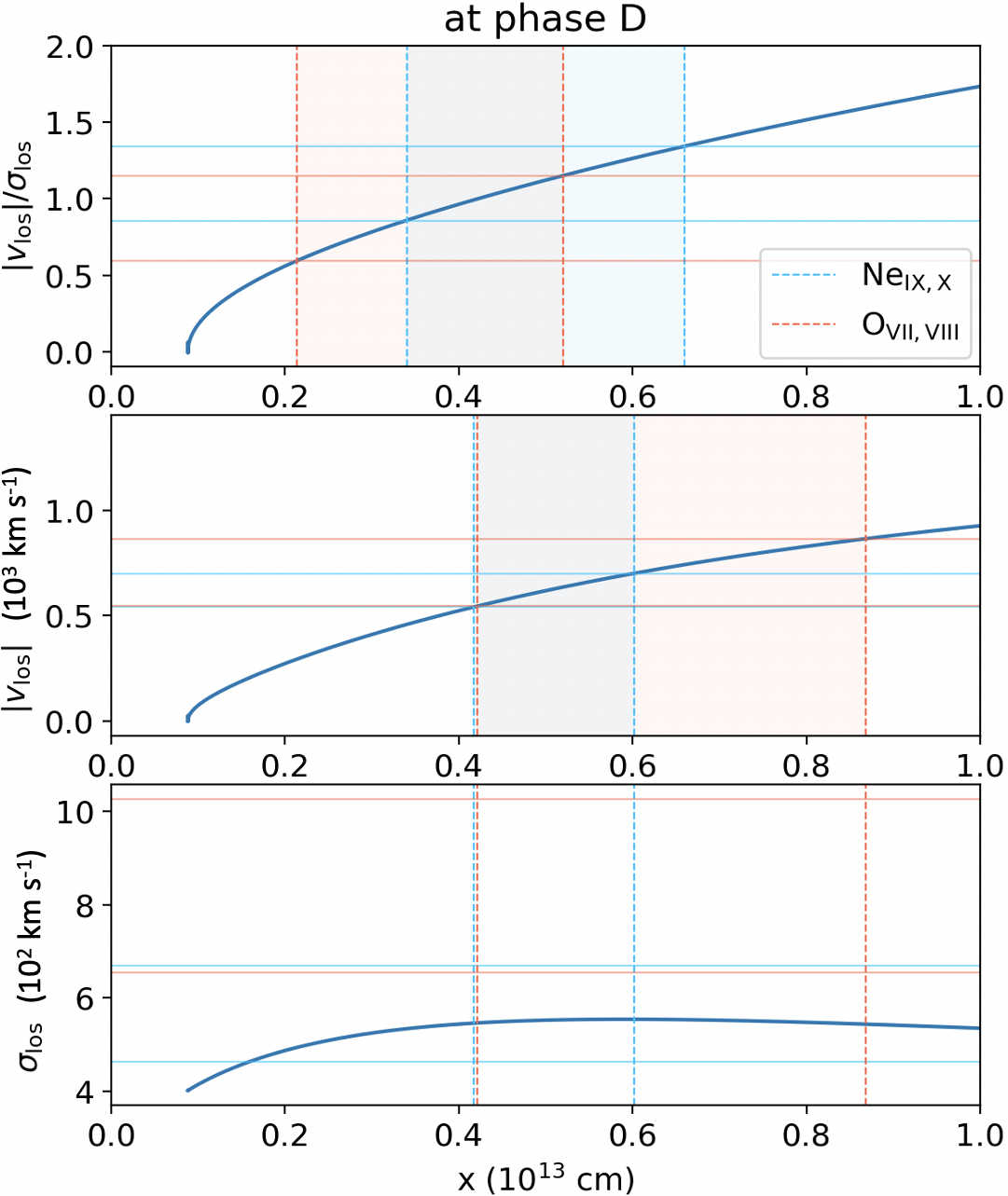}
  \end{minipage}
  \hfill
  \begin{minipage}{0.33\textwidth}
    \caption{$|v_{\rm los}|/\sigma _{\rm los}$ (upper panel), the line-of-sight velocity $|v_{\rm los}|$ (middle panel) and its dispersion $\sigma _{\rm los}$ (lower panel) as a function of the $x$ coordinate. The theoretical curves (blue) are based on our numerical calculations based on \citetalias{10.1093/mnras/stac1289} and Equations~(\ref{eq:vlos}) and (\ref{eq:sigmalos}). 
    The cyan and orange hatches indicate the locations of the K$\alpha$ line-emission site of Ne and O, respectively, which are obtained by intersecting the allowed ranges of the observed $|v_{\rm los}|$/$\sigma _{\rm los}$ and $|v_{\rm los}|$ with the theoretical curves.
    The vertical dashed lines of the lower panels ($\sigma _{\rm los}$) indicate the locations of line-emission sites determined from the middle panel $|v_{\rm los}|$. 
    The allowed ranges of the velocity dispersion are larger than the theoretical ones at the locations of the line-emission sites at earlier phases K (0.816) and A (0.912) for both Ne and O.
    }
    \label{fig:location}
    \end{minipage}
\end{figure*}

In this Section, we compare the observed line-of-sight velocities of the Ne and O emission lines with those calculated in the previous section to identify the locations of these line-emission sites.
Figure~\ref{fig:location} shows the profiles of $|v_{\rm los}|$ (middle panel) and $\sigma _{\rm los}$ (bottom panel) calculated using Equations~(\ref{eq:vlos}) and (\ref{eq:sigmalos}).
We also draw the horizontal lines representing the allowed ranges of $|v_{\rm los}|/\sigma _{\rm los}$, $|v_{\rm los}|$, and $\sigma _{\rm los}$ 
measure using the \ion{Ne}{ix, x} and \ion{O}{vii, viii} lines and determine the ranges $x$ from $|v_{\rm los}|/\sigma_{\rm los}$ and $|v_{\rm los}|$. The vertical lines in the $\sigma_{\rm los}$ panels are identical to those in the $|v_{\rm los}|$ panels.
The theoretical curve and data of Ne in the $|v_{\rm los}|/\sigma _{\rm los}$ panel are 
taken from \citetalias{10.1093/mnras/stac1289}. 
A similar analysis of the Fe lines do not yield restrictive results; hence, hereafter, we concentrate on the Ne and O data. 

The $x$ coordinate of the Ne line-emission site determined with $|v_{\rm los}|$ (middle panels) is more distant from the stagnation point than that determined with $|v_{\rm los}|/\sigma _{\rm los}$ (top panels) at earlier phases K (0.816), A (0.912), and L (0.935). 
Simultaneously, the allowed range of the observed $\sigma_{\rm obs}$
is larger than the theoretical $\sigma_{\rm los}$. 
This implies that there is an additional factor that enhances the velocity dispersion above the calculated $\sigma_{\rm los}$.
Possible additional components for enhancing the velocity dispersion are summarised in Equation~(23) of \citetalias{10.1093/mnras/stac1289}. 
In this equation, $\sigma_{\rm lam}$ is 
equal to $\sigma_{\rm los}$ [Equation~(\ref{eq:sigmalos})] in the present study.
Furthermore, the velocity dispersion associated with the thermal motion of \ion{Ne}{x} ($\sigma_{\rm th}$) is negligible \citepalias{10.1093/mnras/stac1289}.
$\sigma_\perp$, 
originating from the divergence of the plasma while it flows along the shock cone, is expected to be small compared with $\sigma_{\rm lam}$ because the diverging angle of the plasma flow 
\citep[$\simeq 2(\beta - \alpha)$ in Fig.~2 of][for example]{1992ApJ...389..635U}
must be smaller than the opening angle of the shock cone \citep[$= \alpha$ in Fig.~2 of][for example]{1992ApJ...389..635U}.

In addition to $\sigma_{\rm lam}$ and $\sigma_{\rm turb}$ (the velocity dispersion of the turbulence) in \citepalias{10.1093/mnras/stac1289}, we must newly take into consideration the spatial extent of the Ne line-emission site along the shock cone.
If the extent
is sufficiently large, the variation in $|v_{\rm los}|$ along the shock cone, which we denote hereafter as $\sigma_{\rm vlos}$, may not be negligible.
Consequently, Equation~(23) given in \citetalias{10.1093/mnras/stac1289} can now be written as,
\begin{equation}
\sigma_{\rm obs}^2 = \sigma_{\rm los}^2 + \sigma_{\rm turb}^2 + \sigma_{\rm vlos}^2
\label{eq:sigmaobs}
\end{equation}
In summary, the observed velocity-dispersion enhancement is attributed to the turbulence and/or to the variation in the line-of-sight velocity along the shock cone.
In Sections 4.3 and 4.4, we consider these possibilities in detail.

As described in Section 4.1, the locations of the line-emission sites should be measured using $|v_{\rm los}|$ rather than using $|v_{\rm los}|/\sigma _{\rm los}$ because the observed $\sigma_{\rm obs}$ is now found larger than the calculated $\sigma_{\rm los}$ expected based on the laminar flow.
Table~\ref{tab:NeOlocation} summarises the line-emission site locations, updated with $|v_{\rm los}|$ 
using the middle panels of Fig.~\ref{fig:location}. 
\begin{table*}
  \centering
  \caption{Locations of the \ion{Ne}{ix,x} and \ion{O}{vii,viii} line-emission site. $r$ is the distance from the stagnation point.\label{tab:NeOlocation}}
    {\tabcolsep = 8pt
  \begin{tabular}{lccccccc} \hline 
    	&\multicolumn{3}{c}{\ion{Ne}{ix, x}}&&\multicolumn{3}{c}{\ion{O}{Vii, viii}}\\ \cline{2-4} \cline{6-8}
    Phase & $x$ ($10^{13}$ cm) & $y$ ($10^{13}$ cm) & $r$ ($10^{13}$ cm) && $x$ ($10^{13}$ cm) & $y$ ($10^{13}$ cm) & $r$  ($10^{13}$ cm)\\
	\hline
	K (0.816)& 4.8 - 6.8& 8.9 - 10.5 & 10.1 - 12.6 &&5.1 - 7.5&9.2 - 11.0&10.5 - 13.4\\
	A (0.912)& 2.5 - 3.4& 5.1 - 6.0 &5.7 - 6.9&&2.7 - 3.8&5.3 - 6.3&6.0 - 7.4\\
	L (0.935)& 2.9 - 3.9&5.0 - 5.8&5.8 - 7.0&&2.6 - 3.5&4.7 - 5.5&5.4 - 6.6\\
	B (0.968)& 1.6 - 2.1&2.9 - 3.2&3.3 - 3.9&&1.6 - 2.3&2.9 - 3.4&3.3 - 4.1\\
	D (0.987)& 0.4 - 0.6&1.0 - 1.2&1.0 - 1.3&&0.4 - 0.9&1.0 - 1.5&1.1 - 1.7\\
   \hline
  \end{tabular}}
\end{table*}
The distance $r$ from the stagnation point for both the Ne and O line-emission sites range from $1\times 10^{13}$~cm at phase D (0.987) to $13\times 10^{13}$~cm at phase K (0.816). 
These locations correspond to the spatial centroids of the line-emission sites, and their spatial extents will be considered in Sections 4.3.

At later phases B (0.968) and D (0.987), the $x$ coordinates from $|v_{\rm los}|/\sigma _{\rm los}$ and $|v_{\rm los}|$ are consistent both for Ne and O, and the allowed ranges of the observed $\sigma_{\rm los}$ shown in the lower panel overlaps with the theoretical curve [except for O at phase D (0.987)]. This implies that no significant turbulence is detected at these phases.

\subsection{Spatial extent of the line-emission sites along the shock cone}

In this Section, we explore the spatial extents of the Ne and O line-emission sites 
along the shock cone.
Now that we know the location, the temperature, and the flow velocity of the O and Ne line-emission sites, and we can calculate the densities there from the local ram-pressure balance of the stellar winds, we can
calculate the thermal energy that the line-emission plasmas possess, evaluate their cooling time, and finally obtain their spatial extent along the shock cone as a product of plasma flow velocity and cooling time. 
We do not intend to derive any strict solution of the shock cone plasma but just calculate radiative cooling of the plasma based on the observed quantities and elementary fluid mechanics.  
In this Section, the spatial extent of 
the Ne line-emission site at phase K (0.816), as an example, is discussed in detail. 
The results of Ne and O in all phases are summarised in Table~\ref{tab:ring_val1} and \ref{tab:ring_val2}.

\subsubsection{Thermal energy and cooling rate of the line-emission plasma}

$|v_{\rm los}|$ and $\sigma _{\rm los}$ of the Ne line-emission plasma are determined primarily by the \ion{Ne}{x} K$\alpha$ line, because it is more intense and narrower than \ion{Ne}{ix} \citepalias[Fig.~3 of][]{10.1093/mnras/stac1289}.
We therefore first explore the temperature of the plasma that radiates \ion{Ne}{x} K$\alpha$ based on atomic data.
We refer to \citet{1985A&AS...62..197M} for the so-called ‘cooling coefficient’ of the \ion{Ne}{x} K$\alpha$ line $\Lambda_\ion{Ne}{x}(T)$ (photons~cm$^3$~s$^{-1}$).
We then multiply this by the square of the plasma particle number density $n^2(T)$ to obtain the emissivity $\varepsilon_\ion{Ne}{x}(T)$ (photons~cm$^{-3}$~s$^{-1}$). 
For the density $n (T)$, we utilise the fact that the post-shock plasma flow in the shock cone is isobaric, that is,
\begin{equation}
    n(T_{\rm Ne}) = \frac{T_{\rm S}}{T_{\rm Ne}}n_{\rm S},
    \label{eq:IsobaricFlow}
\end{equation}
where $T_{\rm S}$ and $T_{\rm Ne}$ are the temperatures of the plasma at the stagnation point and the Ne line-emission site, which is 3.5~keV \citep{2015PASJ...67..121S} and 0.453~keV at phase K (0.816) \citepalias[Table~5 of][]{10.1093/mnras/stac1289}, respectively.
$n_{\rm S}$ is the plasma density at the stagnation point, which is equal to 4$\times$$(n_{\rm rw}+n_{\rm o})/2$, where $n_{\rm wr}$ and $n_{\rm o}$ are the densities of the winds of the WR star and the O star (Table~\ref{tab:par2}). 
The calculated $n(T_{\rm Ne})$ values are listed as $n(T_{\rm Ne/O})$ in Table~\ref{tab:ring_val1}.
As the other way, we can estimate the density of the Ne line-emission site under the assumption of a local pressure balance with the WR wind using the location of the Ne line-emission site updated in the previous Section.
Detailed calculation is provided in APPENDIX B1.
The density thus obtained is also summarised in Table~\ref{tab:ring_val1} as $n_{\rm Ne,\,loc}$.
These two densities coincide each other within a factor of 2.

\begin{figure}
	\includegraphics[width=\columnwidth]{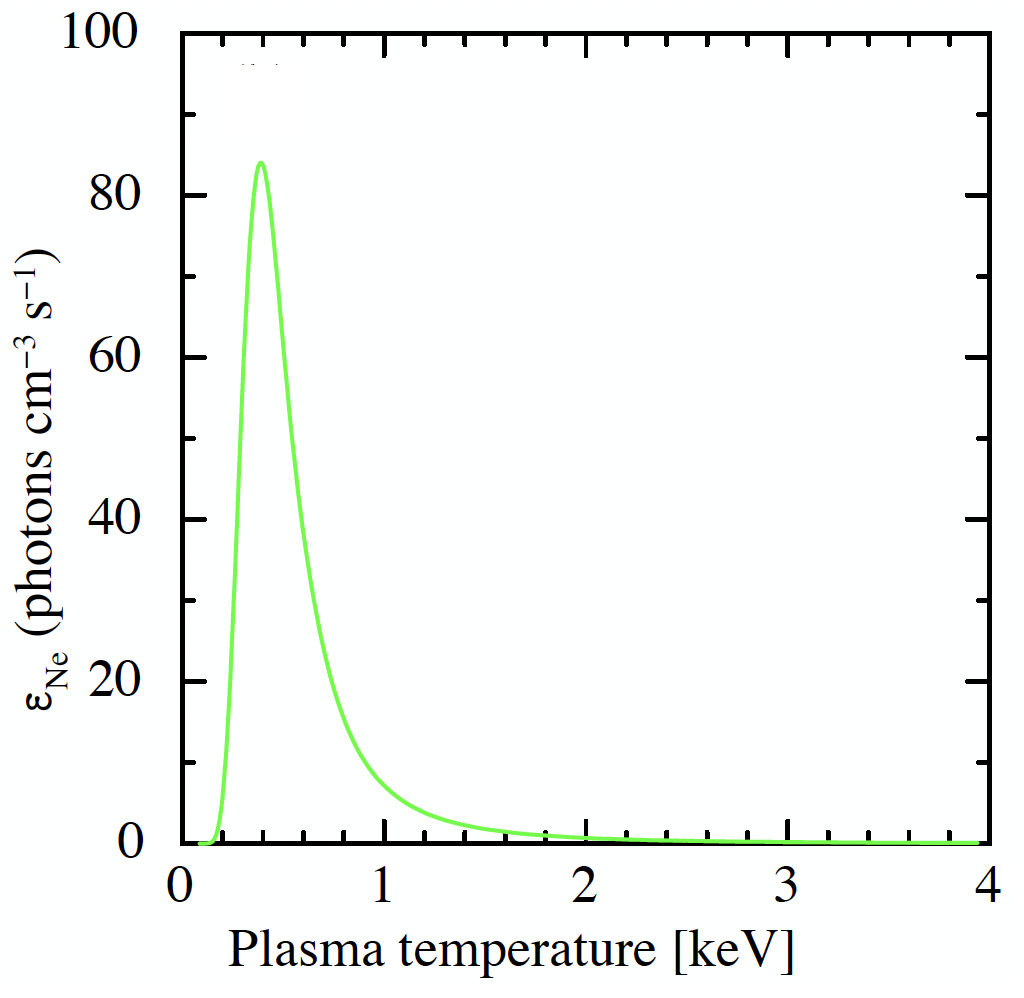}
    \caption{Profile of $\varepsilon_\ion{Ne}{x}(T)$ obtained with $\Lambda_\ion{Ne}{x}(T)$ tabulated in \citet{1985A&AS...62..197M} multiplied by $n^2(T)$, where $n(T)$ is obtained with the temperature and density at the stagnation point $kT_{\rm S} = 3.5$~keV and $n_{\rm S} = 4n_{\rm wr} = 9.2\times10^{6}$ cm$^{-3}$ through $n(T) = (T_{\rm S}/T) n_{\rm S}$ [equation~(\ref{eq:IsobaricFlow})].}
    \label{fig:varepsilon}
\end{figure}

The profile of $\varepsilon_\ion{Ne}{x}(T)$ [$= \Lambda_{\rm\ion{Ne}{x}}(T) n^2(T)$] is shown in Fig. \ref{fig:varepsilon}.
The \ion{Ne}{x} K$\alpha$ line emanates from a region with a temperature of $\lesssim$1 keV. This excludes the stagnation point ($kT_{\rm S} = 3.5$~keV) as a \ion{Ne}{x} K$\alpha$ line-emission site, and the observed \ion{Ne}{x} K$\alpha$ line is radiated from a ‘ring’ on the shock cone.
Based on this figure, we find that the 1$\sigma$ full width of the temperature
of $\varepsilon_\ion{Ne}{x}(T)$ is $kT$ = 0.29-0.61 keV, or $k\Delta T$ = 0.32 keV. This means that the 68\% ($= \pm$1$\sigma$) of the Ne line photons are emitted while the shock cone plasma is cooled from the temperature 0.61~keV to 0.29~keV.
The centroid of this temperature range is $kT = 0.45$~keV
which coincides with the Ne line-emission site temperature at phase K (0.816) \citepalias[Table~5 of][]{10.1093/mnras/stac1289}.
\begin{table*}
  \centering
  \caption{Values of the parameters used in the calculation to obtain the spatial extent of the Ne and O line-emission sites along the shock cone. 
  $n(T_{\rm Ne/O})$ is the particle number density at the Ne or O line-emission site under the assumption of isobaric plasma flow,
  $\theta _{\rm wr,ring}$ is the angle between the WR wind orientation and the tangent of the shock cone at the Ne or O line-emission site (= $\theta_{\rm wr}$ in Fig.~\ref{fig:zahyou}), 
  $p_1$ is the ram-pressure of the WR wind at the Ne or O line-emission site,
  $n_{\rm Ne/O,\,loc}$ is the particle number density at the Ne or O line-emission site under the assumption of a local pressure balance with the WR wind, 
  $\varepsilon _{\rm Fe}$ is the emissivity due to iron emission lines (including that of Si and S lines for O), 
  $\varepsilon _{\rm brems}$ is the emissivity of the thermal bremsstrahlung,  
  $\tau$ is the cooling time of the Ne and O line-emission plasma, 
  $V_{\rm Ne/O}$ is the local plasma flow velocity at the Ne or O line-emission site [equation~(\ref{eq:V})], 
  $w_{\rm ring}$ is the spatial extent of the Ne or O line-emission site along the shock cone, 
  and $l_{\rm ring}$ is the distance of the Ne or O line-emission site from the stagnation point along the shock cone. \label{tab:ring_val1}}
   {\tabcolsep = 3pt
  \begin{tabular}{lccccccccccc} \hline
     & Phase& $n(T_{\rm Ne/O})$ & $\theta _{\rm wr,ring}$& $p_1$ & $n_{\rm Ne/O,\,loc}$ & $\varepsilon _{\rm Fe}$&$\varepsilon _{\rm brems}$ & $\tau$ &$V_{\rm Ne/O}$ &$w_{\rm ring}$& $l_{\rm ring}$ \\
     & & (cm$^{-3}$) & (degrees)& (dyne cm$^{-2}$) & (cm$^{-3}$)& (erg cm$^{-3}$ s$^{-1}$)&(erg cm$^{-3}$ s$^{-1}$) &(sec) &(10$^{3}$ km~s$^{-1}$)& (cm) &(cm)\\ \hline
    & K (0.816) &2.2$\times 10^8$ & 21.8&3.3$\times 10^{-2}$ &9.0$\times 10^7$ &1.3$\times 10^{-6}$&4.6$\times 10^{-8}$ &3.4$\times 10^{4}$&1.83&6.2$\times 10^{12}$&1.2$\times 10^{14}$\\
     & A (0.912) &5.9$\times 10^8$ & 25.0&1.3$\times 10^{-1}$ &3.9$\times 10^8$ &2.9$\times 10^{-5}$&8.2$\times 10^{-7}$&6.5$\times 10^{3}$&1.73&1.1$\times 10^{12}$&1.7$\times 10^{13}$\\
    \ion{O}{vii, viii} & L (0.935) & 8.0$\times 10^8$& 22.3& 1.5$\times 10^{-1}$&4.0$\times 10^8$ &2.6$\times 10^{-5}$&9.2$\times 10^{-7}$&7.5$\times 10^{3}$&1.78&1.3$\times 10^{12}$&1.5$\times 10^{13}$\\
     & B (0.968) & 1.6$\times 10^{10}$& 21.8&3.7$\times 10^{-1}$ &9.8$\times 10^8$ &1.6$\times 10^{-4}$&5.5$\times 10^{-6}$&3.1$\times 10^{3}$&1.77&5.4$\times 10^{11}$&8.2$\times 10^{12}$\\
     & D (0.987) & 9.1$\times 10^9$& 32.6&3.3 &9.6$\times 10^9$ &8.7$\times 10^{-3}$&5.1$\times 10^{-4}$&2.6$\times 10^{2}$&1.49&3.9$\times 10^{10}$&1.4$\times 10^{13}$\\ 
     \hline
     & K (0.816) &1.1$\times 10^8$ & 21.8&4.0$\times 10^{-2}$& 5.5$\times 10^7$ & 4.6$\times 10^{-8}$& 1.1$\times 10^{-8}$& 7.4$\times 10^{5}$&1.80&1.3$\times 10^{14}$&1.2$\times 10^{14}$\\
     & A (0.912) &2.6$\times 10^8$ &27.0&1.6$\times 10^{-1}$& 2.1$\times 10^8$ & 6.2$\times 10^{-7}$& 1.6$\times 10^{-7}$&2.0$\times 10^{5}$&1.70&3.4$\times 10^{13}$&6.6$\times 10^{13}$\\
    \ion{Ne}{ix, x} & L (0.935) & 3.9$\times 10^8$& 20.1 & 1.2$\times 10^{-1}$&1.5$\times 10^8$&3.5$\times 10^{-7}$& 8.8$\times 10^{-8}$&2.7$\times 10^{5}$&1.82&4.9$\times 10^{13}$&6.6$\times 10^{13}$\\
     & B (0.968) &4.4$\times 10^9$ &22.8& 4.1$\times 10^{-1}$& 3.1$\times 10^8$&9.8$\times 10^{-7}$& 4.5$\times 10^{-7}$&1.6$\times 10^{5}$&1.75&2.9$\times 10^{13}$&3.7$\times 10^{13}$\\
     & D (0.987) &3.3$\times 10^9$ &38.7&4.8& 5.1$\times 10^9$&1.0$\times 10^{-3}$ & 1.0$\times 10^{-4}$& 8.4$\times 10^{3}$&1.41&1.3$\times 10^{12}$&1.2$\times 10^{13}$\\
    \hline
  \end{tabular}
  }
\end{table*}

\begin{table*}
  \centering
  \caption{Magnitude of the velocity dispersion components derived from the calculated spatial extent of the Ne or O line-emission region. $|\Delta v_{\rm los}|$ is the range of $|v_{\rm los}|$ within the $w_{\rm ring}$ in Table~\ref{tab:ring_val1} and $\sigma _{v{\rm los}}$ is the dispersion of $|v_{\rm los}|$ (= $|\Delta v_{\rm los}|/2$). $\sigma _{\rm turb}$ is the dispersion of the turbulence velocity. \label{tab:ring_val2}}
   {\tabcolsep = 8pt
  \begin{tabular}{lccccc} \hline
     & Phase & $|\Delta v_{\rm los}|$ & $\sigma_{v{\rm los}}$ & $\sqrt{\sigma _{v{\rm los}}^2 + \sigma_{\rm turb}^2}$ & $\sigma_{\rm turb}$ \\
     &       & (km~s$^{-1}$)   & (km~s$^{-1}$) & (km~s$^{-1}$) & (km~s$^{-1}$) \\ \hline
     & K (0.816) & 1125-1150 & 11.3 & 396-660 & 396-660\\
     & A (0.912) & 1088-1121 & 16.6 & 160-484 & 160-484\\
    \ion{O}{vii, viii} 
     & L (0.935) & 1167-1212 & 22.6 & $<$409   & $<$409 \\
     & B (0.968) & 1176-1210 & 17.0 & $<$373   & $<$373 \\
     & D (0.987) & 733-734 & 0.7 & 363-869 & 363-869 \\ 
     \hline
     & K (0.816) & 704-1286 & 291.3 & 386-622 & 253-550 \\
     & A (0.912) & 901-1172 & 135.5 & 364-644 & 338-630 \\
    \ion{Ne}{ix, x} 
     & L (0.935) & 1010-1390 & 190.0 & 295-533 & 226-498 \\
     & B (0.968) & 905-1323 & 208.8 & $<$330   & $<$256 \\
     & D (0.987) & 605-650 & 22.5 & $<$376   & $<$375 \\
    \hline
  \end{tabular}
  }
\end{table*}

Next, we consider the emissivity $\varepsilon (T)$ of the plasma.
According to \citet{1993ApJ...418L..25G}, $\varepsilon (T)$ around the temperature of the Ne line-emission site ($kT_{\rm Ne}$ = 0.453~keV) is the sum of line emission and continuum emission. 
Of them, the line emission is dominated by that from Fe. 
As for the continuum, thermal bremsstrahlung is the major cooling process.
Therefore we can write $\varepsilon = \varepsilon_{\rm Fe} + \varepsilon_{\rm brems}$.
Note that \citet{1993ApJ...418L..25G} assume plasma of the solar abundance \citep{1973asqu.book.....A}; hence, we must accommodate it to the case of WR140 \citep[][Table~3]{2015PASJ...67..121S}.
Detailed calculation for this is presented in APPENDIX B2.
The resultant $\varepsilon_{\rm Fe}$ and $\varepsilon_{\rm brems}$ values are summarised in Table~\ref{tab:ring_val1}. Approximately, $\varepsilon_{\rm Fe}$ is larger than $\varepsilon_{\rm brems}$ by a factor of a few for Ne and an order of magnitude for O.

\subsubsection{Spatial extent of the line-emission sites and their line-of-sight velocity dispersion}

\begin{figure}
	\includegraphics[width=\columnwidth]{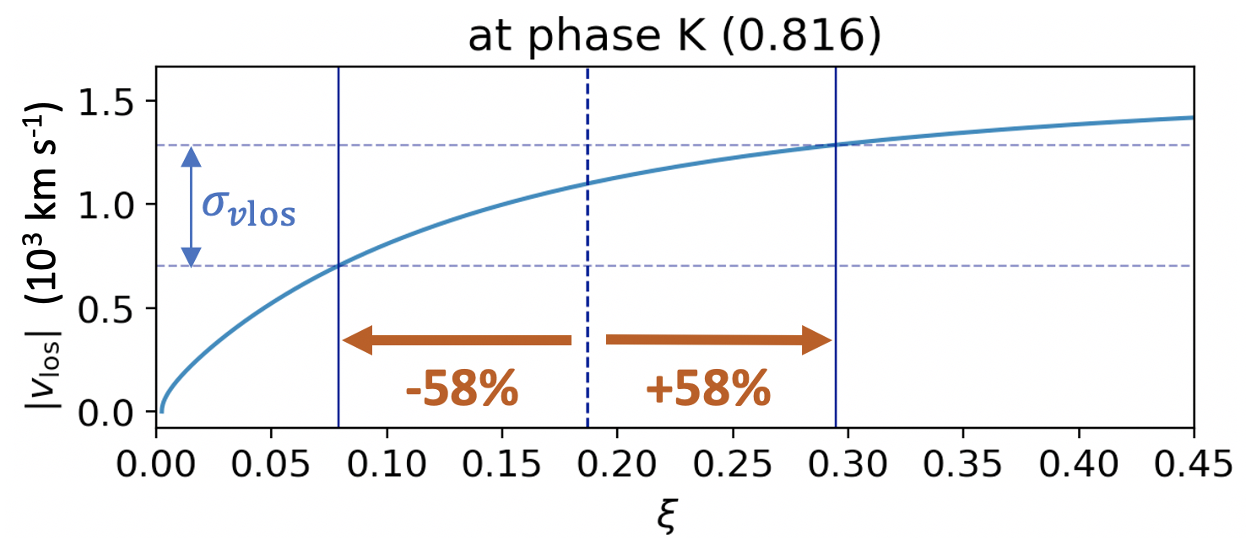}
    \caption{$\sigma_{v{\rm los}}$ estimated with the spatial extent of the Ne line-emission plasma $w_{\rm ring}$ using the theoretical curve of $\sigma_{\rm los}$ (same as the middle panel of Fig. \ref{fig:location}) at phase K(0.816).
    The Ne line-emission plasma has the width $\pm$58 per cent of $l_{\rm ring}$} from the centre of gravity of the line-emission site.
    \label{fig:sigvlos}
\end{figure}

As described in Section 4.3.1, the thermal energy lost from the shock cone plasma while it radiates the \ion{Ne}{x} K$\alpha$ line is $\Delta E=(3/2)n_{\rm Ne,\,loc}k\Delta T$, where $k\Delta T = 0.32$~keV, 
Using this $\Delta E$ and the emissivity $\varepsilon$ (= $\varepsilon_{\rm Fe}+\varepsilon_{\rm brems}$),
the cooling time $\tau$ is given by $\tau=\Delta E/\varepsilon$. 
The Ne line emission takes place from the ring with a width of $w_{\rm ring} = V_{\rm Ne}\tau$ along the shock cone, where $V_{\rm Ne}$ is the local velocity of the plasma at the Ne line-emission site, calculated according to Equations~(\ref{eq:M_dot}) and (\ref{eq:V}).
The values of $\tau$, $V_{\rm Ne}$ and $w_{\rm ring}$ are listed in Table~\ref{tab:ring_val1}. 
$l_{\rm ring}$ tabulated in the last column is the distance from the stagnation point to the Ne line-emission site along the shock cone.
At phase K (0.816), $w_{\rm ring} = 1.3\times 10^{14}$~cm is centered at the spatial centroid of the Ne line-emission site, and is as large as $\pm$58 per cent of $l_{\rm ring}$ (see Fig.~\ref{fig:sigvlos}). This fraction at phase K (0.816) is the largest among all phases for Ne, and it is even smaller for O, 10 per cent at phase L (0.935), and 0.3 per cent at phase D (0.987). Note that, in  \citetalias{10.1093/mnras/stac1289}, we estimated $n_{\rm e} \sim 10^{12}$~cm$^{-3}$ at phase K (0.816). If this large density is confirmed, then $w_{\rm ring}$ will be much smaller than that estimated here.

The variation in the line-of-sight velocity $v_{\rm los}$ within this $w_{\rm ring}$ (hereafter referred to as $|\Delta v_{\rm los}|$) is obtained using the $v_{\rm los}$ curve shown Fig.~\ref{fig:sigvlos} and 
is summarised in Table~\ref{tab:ring_val2} as $\sigma_{\rm vlos}$.
From Fig.~\ref{fig:sigvlos}, $|\Delta v_{\rm los}|$ = (0.70-1.29)$\times 10^3$ km~s$^{-1}$ at phase K (0.816).
This width should be regarded as $\pm$1$\sigma$ of the $v_{\rm los}$ distribution of the entire Ne line emission, namely $\sigma_{\rm vlos} = |\Delta v_{\rm los}|/2$,
because the energy loss $k\Delta T = 0.32$~keV (Section 4.3.1) covers 68 per cent of the $\varepsilon_{\rm\ion{Ne}{x}}(T)$ distribution (Fig.~\ref{fig:varepsilon}).
Consequently, the velocity dispersion associated with the $v_{\rm los}$ variation is 
$\sigma _{v{\rm los}}$ = 291 km~s$^{-1}$ for Ne at phase K (0.816). 
The values of $\sigma _{v{\rm los}}$ for Ne and O at all phases are summarised in Table~\ref{tab:ring_val2}.
They are of order 10-100 km~s$^{-1}$.

We remark that temperature range $k\Delta T = 0.32$ keV is evaluated from the profile of $\varepsilon_\ion{Ne}{x}(T)$ (Fig. \ref{fig:varepsilon}). 
To determine the real temperature range, $\varepsilon_\ion{Ne}{x}(T)$ should be further multiplied by the plasma volume.
Since the plasma flow that cools along the shock cone has a larger volume at lower temperatures, the resultant temperature range is smaller than that shown in Fig.~\ref{fig:varepsilon} as the lower bound of $\varepsilon_\ion{Ne}{x}(T)$ has a sharp cutoff owing of the recombination from \ion{Ne}{x} to \ion{Ne}{ix}. Strictly speaking, the temperature range $k\Delta T = 0.32$ keV should be regarded as the upper limit.

Finally, for evaluating $w_{\rm ring}$ of O K$\alpha$ lines, we have took into account the plasma cooling not only by the Fe lines but also by Si and S lines in a similar manner to Equation~(\ref{eq:varepsilon_Fe}), because the temperature of the \ion{O}{vii,viii} line-emission site is lower ($\sim$0.2~keV). At such temperatures, Si and S cooling work in addition to Fe \citep{1993ApJ...418L..25G}.

\subsection{Evaluation of the turbulent velocity}

In this section, we consider the origin of the excess velocity dispersion detected in Section 4.2. As already discussed there, the excess is expressed with Equation~(\ref{eq:sigmaobs}) as
\begin{equation}
\sqrt{\sigma _{\rm obs}^2 - \sigma _{\rm los}^2}
= \sqrt{\sigma _{\rm turb}^2 + \sigma_{\rm vlos}^2},
\label{eq:sigma_turb}
\end{equation}
where $\sigma_{\rm obs}$ is summarized in Table 5 of \citetalias{10.1093/mnras/stac1289} for Ne and Table~\ref{tab:O78bestfit} for O. $\sigma_{\rm los}$ is found in Fig.~\ref{fig:location}.

Magnitudes of the quantities that appear in Equation~(\ref{eq:sigma_turb}) are summarized in Table~\ref{tab:ring_val2}.
$\sqrt{\sigma_{\rm turb}^2 + \sigma_{\rm vlos}^2}$ is of order 100 km~s$^{-1}$, whereas $\sigma_{\rm vlos}$ is in general smaller than this.
Consequently, we believe that the observed excess of the velocity dispersion 
that cannot be explained with the $v_{\rm los}$ distribution should be attributed to turbulence.
The turbulence velocity dispersion, which is listed in Table~\ref{tab:ring_val2},
is detected at the three early phases K (0.816), A (0.912), and L (0.935) for Ne, whereas 
only the upper limit is obtained at the latter two phases: B (0.968) and D (0.987). 
A similar tendency is also found for O, where turbulence is significantly detected at phases K (0.816) and A (0.912), whereas at the later phases, $\sigma_{\rm turb}$ is an upper limit, except for the last phase D (0.987).
The resultant $\sigma_{\rm turb}$ values of O are summarised in Table \ref{tab:ring_val2}. 
The magnitude of $\sigma_{\rm turb}$ is generally in the order of 100 km~s$^{-1}$ for both Ne and O.

Note that the two earlier orbital phases in which we detect a statistically significant $\sigma_{\rm turb}$ coincide with the phases where extraordinarily high plasma density of up to $\sim$10$^{12}$~cm$^{-3}$ is detected with the He-like triplet of \ion{Ne}{ix} K$\alpha$ line \citepalias{10.1093/mnras/stac1289}. 
Such a high density may be a result of turbulence.

\begin{figure}
	\includegraphics[width=\columnwidth]{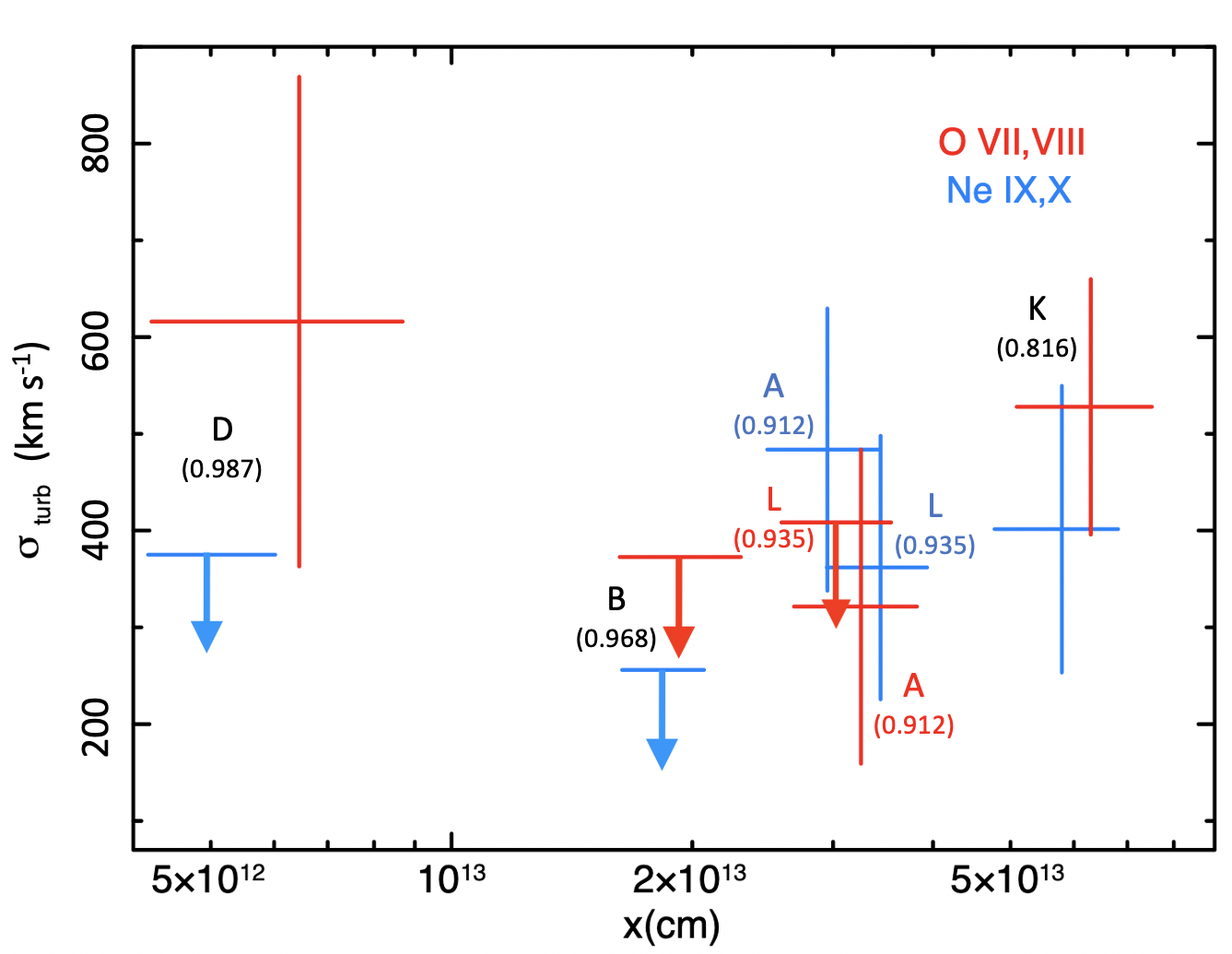}
   \caption{Turbulence velocity dispersion $\sigma _{\rm turb}$ as a function of $x$. 
    }
    \label{fig:x_turb}
\end{figure}

Fig.~\ref{fig:x_turb} shows the plots of $\sigma _{\rm turb}$ listed in Table \ref{tab:ring_val2} as functions of the $x$ coordinate measured from the stagnation point. $\sigma _{\rm turb}$ appears to increase with the distance from the stagnation point. This may indicate growing turbulence in the hot-shocked plasma as it flows along the shock cone. However, as shown in this figure, this trend is not statistically significant. 
Future studies on high accuracy measurements are required to conclude whether this trend is real or not.

\begin{figure}
	\includegraphics[width=\columnwidth]{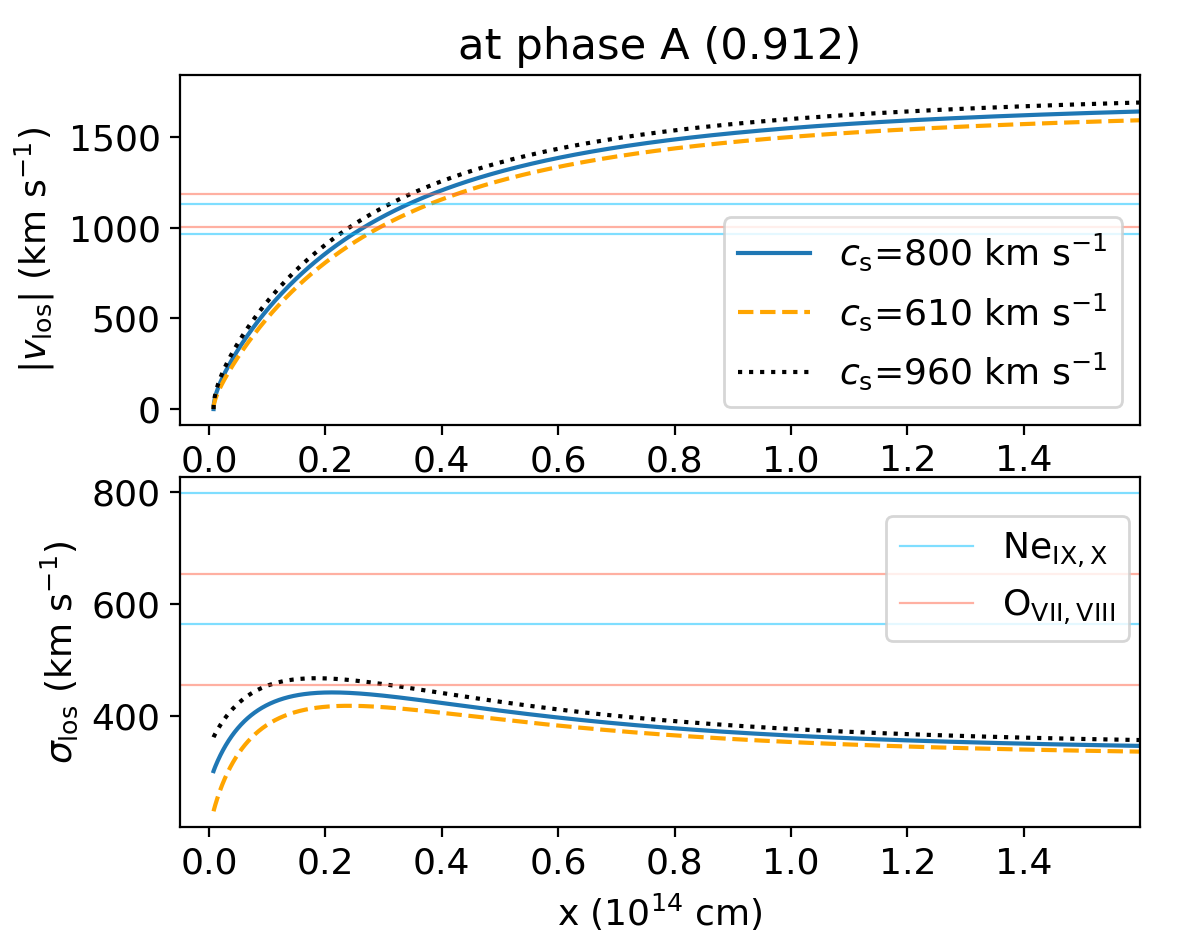}
    \caption{Comparison of the profiles of the line-of-sight velocity $|v_{\rm los}|$ (the upper panel) and its dispersion $\sigma_{\rm los}$ (the lower panel) at phase A (0.912) with initial speeds (= $c_{\rm s}$) = 610 km~s$^{-1}$ (orange, pure WR star composition), 800 km~s$^{-1}$ (blue, averaged composition of the WR- and O stars) and 960 km~s$^{-1}$ (black, pure O star composition). 
    }
    \label{fig:v_s_c}
\end{figure}

Finally, as predicted in Section 4.1.1, we examine the possible uncertainty of the theoretical curves $|v_{\rm los}|$ and $\sigma_{\rm los}$ associated with the uncertainty of the initial velocity at the stagnation point. 
Our claim of turbulence detection is entirely based on the calculated profiles of $\sigma_{\rm los}$ and $|v_{\rm los}|$. Hence, if their uncertainty is too large, 
we would not be able detect the turbulence. 
We calculate $|v_{\rm los}|$ and $\sigma _{\rm los}$ using the initial speeds (= $c_{\rm s}$) = 610 and 960 km~s$^{-1}$,which are the values of pure WR star wind and pure O star wind, respectively. 
Fig. \ref{fig:v_s_c} shows these results together with the case of the initial speeds = 800 km~s$^{-1}$ at phase A (0.912), for example. 
The three $\sigma_{\rm los}$ curves (lower panel) converge at the right end of the $x$ coordinate, whereas the upper panel curves $|v_{\rm los}|$ differ slightly, even at the right end of $x$. 
However, the difference is approximately 6 per cent at full amplitude. Hence, we conclude that the difference in the initial speed seldom affects the characterisation of the plasma flow. 

\subsection{Limitation of our approach and prospect for future study}

In Section 4.3, we have derived the spatial extent of the Ne and O line-emission sites based on the simple radiative cooling calculation. We employ the observed plasma temperatures. The densities are calculated from the stellar wind parameters. The radiative cooling efficiency relies on atomic physics. Although these evaluations are accurate enough for the order-of-magnitude estimation we made on the spatial extent, there remain some uncertainties.
\citet{1992ApJ...389..635U}, for example, presented a "stratified" shock cone plasma model in which the stellar winds keep flowing along their own stream lines even after experiencing the shock \citep[see Fig.~3 of][]{1992ApJ...389..635U}. In such a case, the plasma cooling should be considered for each stream line independently, and the resultant temperature distribution is a superposition of the temperature distributions along the stream lines. Nevertheless, we believe our discussion on the spatial extent of the plasma and the turbulence made an important step forward for understanding the physical state of the colliding stellar wind plasma, since the evaluations are mades based on the simple but clear assumptions.

Our final goal is to understand the physical state of the colliding wind shock plasma, namely to derive distribution of physical quantities, such as the temperature, the density and the velocity of the plasma along the shock cone. For this to be realized, obviously we need full simulation of the wind collision from the theoretical side.
From the observational side, on the other hand, we can achieve our purpose if we can do what we did for the Ne K$\alpha$ emission lines to other abundant metals such as O, Mg, Si, S and Fe. 
This can be done with the {\it Resolve} instrument \citep[X-ray micro-calorimeter,][]{2022HEAD...1920301K,2022SPIE12181E..1SI} onboard the XRISM observatory \citep{2020SPIE11444E..22T} launched in 2023 September.

\section{Conclusion}

We analyse the high-resolution X-ray spectra of the WR+O binary WR140 observed using the RGS onboard {\it XMM-Newton} from May 2008 to June 2016. 
High-quality spectra are obtained when the O star is in front of the WR star.
Following the analysis method for the Ne K$\alpha$ emission lines reported in \citetalias{10.1093/mnras/stac1289}, 
we find that the line-of-sight velocity of \ion{O}{vii,viii} ranges from $-$700 to $-$1200 km~s$^{-1}$, 
and its dispersion ranges from 400 to 800 km~s$^{-1}$,
respectively, and those of \ion{Fe}{xvii,xviii,xx} from $-$800 to $-$1400 km~s$^{-1}$, and from 500 to 1100 km~s$^{-1}$,
respectively. 
These values are approximately the same as those obtained for the Ne emission lines.
From the O, Fe, and Ne emission lines, we confirm that the observed $|v_{\rm los}|$ and $\sigma _{\rm obs}$ are largest and smallest, respectively, between phases B (0.968) and L (0.935), where the inferior conjunction of the O star occurs.
This behaviour of the observed velocities is consistent with the picture in which the colliding wind plasma flows along the shock cone.

We perform a density diagnosis using the intensity ratios of the He-like triplet components of O.
However, we have imposed only upper limits of $n_{\rm e} \lesssim$10$^{10}$-10$^{12}$ cm$^{-3}$ due to statistical limitations 
and uncertainty of the amount of EUV radiation from the O star.
We also attempt to estimate $n_{\rm e}$ using the intensity ratios 17.10 {\AA}, 17.05 {\AA} and 16.78 {\AA} of \ion{Fe}{xvii} \citep{2001ApJ...560..992M}. However, we are not able to obtain any constraints owing to poor statistics and weakness of the lines.

We adopt $v_{\rm los}$ as a more reliable measure of the locations of line emission regions than $|v_{\rm los}|$/$\sigma_{\rm los}$.
We calculate $v_{\rm los}$ using the plasma flow velocity $V$ [Equations~(\ref{eq:M_dot}) and (\ref{eq:V})].
As a result, we find that the location of the Ne line-emission site measured with $|v_{\rm los}|$ is more distant from the stagnation point than that with $|v_{\rm los}|$/$\sigma _{\rm los}$ at the earlier orbital phases K (0.816), A (0.912), and L (0.935), and the observed velocity dispersion $\sigma_{\rm obs}$ is larger than the calculated $\sigma_{\rm los}$ at these phases. 
We update the distance of the Ne line-emission sites using $|v_{\rm los}|$ to be from 1$\times$10$^{13}$~cm [phase D (0.987)] to 13$\times 10^{13}$~cm [phase K (0.816)] (Table \ref{tab:NeOlocation}). 
The values of the newly measured O line-emission sites are similar.
Based on the observed temperatures, the densities calculated from the stellar wind parameters, and the atomic physics for the radiative cooling efficiency, 
we have found that the Ne and O line-emission regions  
extend along the shock cone by up to $\pm$58 per cent (Ne lines at phase K(0.816) of the distance from the stagnation point along the shock cone.
The variation of $|v_{\rm los}|$ within this ‘ring’ is, however, considerably smaller than $\sigma_{\rm obs}$.
This implies that the excess observed velocity dispersion $\sqrt{\sigma_{\rm obs}^2 - \sigma_{\rm los}^2}$ 
contains the turbulence component. 
We find that the maximum turbulent velocity dispersion $\sigma_{\rm turb}$ of Ne is 340-630 km~s$^{-1}$ at phase A (0.912). A similar maximum $\sigma_{\rm turb}$ of 400-660 km~s$^{-1}$ is also obtained for O at 
phase K (0.816) (Table~\ref{tab:ring_val2}). 
At the later phases B (0.968) and D (0.987), we  
obtain no excess of the velocity dispersion from 
Ne. A similar trend is observed for the O lines. 

Based on the plot of $\sigma_{\rm turb}$ versus the distance from the stagnation point $x$ (Fig.~\ref{fig:x_turb}), we suggest that the turbulence in the hot-shocked plasma increases as the plasma flows along the shock cone. Because of statistical limitations, however, future high quality measurements must be conducted before drawing a conclusion whether this trend is real or not.

\section*{Acknowledgements}
The research conducted in this study was supported by NASA under grant number 80GSFC21M0002.
K.H. was supported by NASA grants 15-NUSTAR215-0026 and 80NSSC19K0690 as well as JPL grant
001287-00001.
This study used the Astrophysics Data System and the HEASARC archive. This research was partially supported by the Ministry of Education, Culture, Sports, Science and Technology
(MEXT) and Grant-in-Aid Nos.19K21886 and 20H00175. 
AFJM is grateful for the financial aid from NSERC (Canada).
CMPR acknowledges support from the National Science Foundation under grant no. AST-1747658. The authors are grateful to Dr Ian Stevens for their useful comments.
Finally, we would like to thank Editage (www.editage.com) for English language editing.

\section*{Data Availability}
The data that support the findings of this study are available in the $XMM-Newton$ science archive at http://nxsa.esac.esa.int/nxsa-web/\#home; reference numbers 0555470701, 0555470801, 0555470901, 0555471001, 0555471101, 0555471201, 0651300301, 0651300401, 0762910301, and 0784130301. These data were reduced and analysed using the following resources available in the public domain: XSPEC, which is a part of HEAsoft (https://heasarc.gsfc.nasa.gov/docs/software/lheasoft/); SAS (https://www.cosmos.esa.int/web/xmm-newton/sas); and AtomDB (http://www.atomdb.org/index.php).

\bibliographystyle{mnras}
\bibliography{WR140_MI}

\appendix
\section{Reason for ignoring the Coriolis force}
 
In this section, we show that the Coriolis force is not strong enough to significantly deviate the shape of the shock cone from axial symmetry at the orbital phases we analyse in this work.
We adopt the coordinate system in which the WR star is fixed at one of the foci of the eliptical orbit of the O star.
The acceleration of the Coriolis force {\boldmath $a$}$_{\rm cor}$ acting on the O star is expressed as,
\begin{equation}
    \mbox{\boldmath $a$}_{\rm cor} = -2 \mbox{\boldmath $\omega$}_{\rm orb}\times\mbox{\boldmath $v$}_{\rm orb},
\label{eq:a_cor_vect}
\end{equation}
where {\boldmath $\omega$}$_{\rm orb}$ is the angular velocity of the O star, whose norm is the time derivative of the true anomary $\theta_{\rm orb}$; namely,
\begin{equation}
\omega _{\rm orb}=\frac{{\rm d}\theta _{\rm orb}}{{\rm d}t},
\label{eq:omega_orb}
\end{equation}
and \mbox{\boldmath $v$}$_{\rm orb}$ denote the velocity of the O star.
\begin{equation}
v_{\rm orb}={\displaystyle \sqrt{\left( \frac{{\rm d}d}{{\rm d}t}\right)^2 + d^2\left(\frac{{\rm d}\theta _{\rm orb}}{{\rm d}t}\right)^2}} = {\displaystyle \sqrt{ \dot d^2 + d^2\omega_{\rm orb}^2}},
\label{eq:v_omega}
\end{equation}
where $d$ denotes the orbital separation at each phase. 
As {\boldmath $\omega$}$_{\rm orb}$ and {\boldmath $v$}$_{\rm orb}$ are mutually perpendicular, 
\begin{equation}
a_{\rm cor}=2\omega _{\rm orb}v_{\rm orb}.
\label{eq:a_cor}
\end{equation}

The quantities that appear in Equation~(\ref{eq:a_cor_vect}) and (\ref{eq:a_cor}) are summarised in Table~\ref{tab:Coriolis} and are calculated as follows:
We employ $\Delta t = 2.5\times 10^4$~s as the time interval to evaluate the derivatives, which corresponds to the orbital phase interval $\Delta\phi_{\rm orb} = 0.0001$, and $\dot d$ and $\omega_{\rm orb}$ are calculated as the increments of $d$ and $\theta_{\rm orb}$ over $\Delta t$. We adopt the orbital parameters of \citet{2011ApJ...742L...1M}. After obtaining $\dot d$, $d$, and $\omega_{\rm orb}$, $v_{\rm orb}$ and $a_{\rm cor}$ are calculated using Equations~\ref{eq:v_omega} and \ref{eq:a_cor}.

To obtain the transverse velocity using acceleration $a_{\rm cor}$, the typical escape time of the shock cone plasma must be calculated, which can be defined as follows:
\begin{equation}
    t_{\rm esc} = \frac{10d}{V_{\rm flow}}.
\end{equation}
We approximate the velocity of the plasma in the shock cone $V_{\rm flow}$ as the terminal velocity of the wind from the O star (= 3000 km~s$^{-1}$).
The resultant transverse velocity $v_{\rm cor}$ obtained using the Coriolis force is summarised in the last column of Table~\ref{tab:Coriolis}.
Even at phase D (0.987), its value is 114 km~s$^{-1}$, which is more than an order of magnitude smaller than $V_{\rm flow} = 3000$ km~s$^{-1}$. The values of $v_{\rm cor}$ at the other phases are even smaller than that at phase D (0.987).
Based on these results, we conclude that the Coriolis force can be neglected when the axial symmetry of the shock cone shape is considered.

\begin{table*}
  \centering
  \caption{ {Transverse velocity of the shock cone plasma $v_{\rm cor}$ gained by the Coriolis force.
  $d$ and $\dot d$ represent the distance between the WR star and the O star and its time derivative, respectively. $\omega _{\rm orb}$ and $v_{\rm orb}$ are the angular and orbital velocities of the O star, respectively. $a_{\rm cor}$ is the acceleration by the Coriolis force, and $v_{\rm cor} = a_{\rm cor}t_{\rm esc}$ is the transverse velocity of the shock cone plasma gained by the Coriolis force over time when the plasma moves over a distance of $10d$. \label{tab:Coriolis}}}
  \begin{tabular}{lcccccc} \hline 
  Phase & $\dot d$ & $d$ & $\omega_{\rm orb}$ & $v_{\rm orb}$ & $a_{\rm cor}$ & $v_{\rm cor} (= a_{\rm cor}t_{\rm esc})$ \\
        & (km~s$^{-1}$) & (10$^{13}$ cm) & ($10^{-8}$ s$^{-1}$) & (km~s$^{-1}$) & (10$^{-5}$ km~s$^{-2}$) & (km~s$^{-1}$) \\
	\hline
	K (0.816)&~~34.9 & 31.02 &~~0.61 &~~39.7 & 0.0484   & 5.00 \\
	A (0.912)&~~57.8 & 20.26 &~~1.42 &~~64.6 & 0.183~~  & 12.4~~\\
	L (0.935)&~~67.5 & 16.66 &~~2.10 &~~76.0 & 0.319~~  & 17.7~~\\
	B (0.968)&~~90.7 & 10.25 &~~5.56 & 107.1 & 1.19~~~~ & 40.7~~\\
	D (0.987)& 114.2 &~~5.40 & 20.05 & 158.1 & 6.34~~~~ & 114~~~~~~\\
   \hline
  \end{tabular}
\end{table*}

\section{Supplemental calculations for the derivation of the spatial extent of the line-emission sites along the shock cone}

In this section, we present some supplemental calculations to support derivation of the spatial extents of the line-emission sites along the shock cone (Section 4.3). 
As an example, the Ne line-emission site at phase K (0.816) is discussed in the remainder of this section. The results for Ne and O at all the phases 
are summarised in Table~\ref{tab:ring_val1}.
 
\subsection{Density at the line-emission sites with local pressure balance}

The local pressure balance can be utilised to calculate the densities of the line-emission sites. 
The pre-shock particle number density $n_{\rm wr}$ at the line-emission ring is calculated as follows: 
\begin{equation}
n_{\rm wr}=\frac{\dot M_{\rm wr}}{4\pi r_{\rm wr}^2\mu _{\rm wr}m_{\rm H}v_{{\rm wr,}\infty}},
\label{eq:n_local}
\end{equation}
where $m_{\rm H} = 1.67\times 10^{-24}$~g, $\dot M_{\rm wr}$ = $2.2\times 10^{-5}$ $M_\odot$ yr$^{-1}$, $v_{{\rm wr,}\infty}=2860$ km~s$^{-1}$, and $\mu _{\rm wr} = 1.52$ are the mass of hydrogen, the mass-loss rate, the terminal velocity, and the mean molecular weight of the WR wind (see Section 4.1.1).
$r_{\rm wr}$ is the distance between the Ne line-emission ring and the WR star (Table~\ref{tab:NeOlocation}).
With all these parameters, we obtain the ram-pressure applied to the shock cone as,
\begin{equation}
p_1 = \mu_{\rm wr} m_{\rm H} n_{\rm wr} (v_{\rm wr,\infty} \sin\theta_{\rm wr,ring})^2,
\label{eq:p_1}
\end{equation}
where the angles $\theta _{\rm wr,ring}$ summarised in Table \ref{tab:ring_val1} is the grazing angle of the WR wind to the surface of the shock cone ($\theta _{\rm wr}$ in Fig.~\ref{fig:zahyou}).
As a result, the density of the Ne line-emission site is
\begin{equation}
    n_{\rm Ne,loc} = \frac{p_1}{kT_{\rm Ne}},
    \label{eq:nNelineloc}
\end{equation}
which yields $5.54\times 10^{7}$ cm$^{-3}$ at phase K (0.816).
$n_{\rm Ne,\,loc}$ values are summarised in Table~\ref{tab:ring_val1} for all the phases, which are in agreement with $n(T_{\rm Ne})$ obtained under the assumption of isobaric flow, within a factor of two for Ne and O.

\subsection{Emissivity of the plasmas at the line-emission sites}

We consider the emissivity $\varepsilon (T) = \varepsilon _{\rm Fe}(T)+\varepsilon _{\rm brems}(T)$ of the 
line-emission plasmas whose metal abundances result from appropriate mixture of the winds from the WR star and O star. As an example, we show the process for Ne line-emission plasma at phase K (0.816).

The cooling coefficient $\Lambda (T)$ is shown in Fig. 1 of \citet{1993ApJ...418L..25G}
However, it is given at the solar abundance (such $\Lambda$ is hereafter denoted as $\Lambda_\odot$).
Hence, we must convert $\Lambda_\odot$ to fit the abundance of the mixed plasma. 
The cooling coefficient in \citet{1993ApJ...418L..25G} is first decomposed into the Fe line and thermal bremsstrahlung parts as,
\begin{equation}
\Lambda_\odot (T) = \Lambda_{\rm Fe,\odot} (T) + \Lambda_{\rm brems,\odot} (T),
\label{eq:coolingcoef}
\end{equation}
where $\Lambda_{\rm Fe,\odot} (T_{\rm Ne}) = 3.2\times 10^{-23}$ erg~cm$^3$~s$^{-1}$ is read using the naked eye in Fig.~1 of \citet{1993ApJ...418L..25G}.
According to them, the emissivity of the Fe lines is defined as,
\begin{equation}
    \varepsilon_{\rm Fe,\odot} (T) = \Lambda_{\rm Fe,\odot} (T)n_{\rm e,\odot}\,n_{\rm p,\odot},
    \label{eq:varepsilon_solar}
\end{equation}
where $n_{\rm e,\odot}$ and $n_{\rm p,\odot}$ are the electron and proton densities, respectively, at the solar abundance \citep{1973asqu.book.....A}.
Using $n_{\rm Fe,\odot} = A_{\rm Fe,Allen}\,n_{\rm p,\odot}$, equation~(\ref{eq:varepsilon_solar}) is written as,
 
\begin{equation}
\varepsilon_{\rm Fe,\odot}(T)=\frac{\Lambda _{\rm Fe,\odot}(T)}{A_{\rm Fe,Allen}}n_{\rm e,\odot}n_{\rm Fe,\odot},
\label{eq:varepsilon_Fe_solar}
\end{equation}
where $n_{\rm Fe,\odot}$ is the number density of Fe at the solar abundance and $A_{\rm Fe,Allen}$ is the solar abundance of iron adopted in \citet{1993ApJ...418L..25G} [= $3.98\times 10^{-5}$ relative to hydrogen \citep{1973asqu.book.....A}].
By converting $n_{\rm e,\odot}$ and $n_{\rm Fe,\odot}$ into those of the mixed plasma, 
$n_{\rm e}$ and $n_{\rm Fe}$, we can obtain the emissivity of the iron lines as,
\begin{equation}
\varepsilon_{\rm Fe}(T)=\frac{\Lambda _{\rm Fe,\odot}(T)}{A_{\rm Fe,Allen}}n_{\rm e}n_{\rm Fe}.
\label{eq:varepsilon_Fe}
\end{equation}
In order to calculate $\varepsilon_{\rm Fe}(T)$, we need to express $n_{\rm e}$ and $n_{\rm Fe}$ in Equation~(\ref{eq:varepsilon_Fe}) with the density of the Ne line-emission region $n_{\rm Ne,loc}$ given in Equation~(\ref{eq:nNelineloc}), which is directly linked with the stellar wind densities (Table~\ref{tab:par2}). The density $n_{\rm Ne,loc}$ is a mixture of that of the WR star and the O star;
\begin{equation}
    n_{\rm Ne,loc} = n_{\rm Ne,loc,wr} + n_{\rm Ne,loc,o},
    \label{eq:Neloc}
\end{equation}
where the relative weight is
\begin{eqnarray}
n_{\rm Ne,loc,wr} &=& \frac{\frac{\dot M_{\rm wr}}{\mu _{\rm wr}}(1-\cos \psi _{\rm wr})}{\frac{\dot M_{\rm wr}}{\mu _{\rm wr}}(1-\cos \psi _{\rm wr})+\frac{\dot M_{\rm o}}{\mu _{\rm o}}(1-\cos \psi _{\rm o})}n_{\rm Ne,loc} \nonumber \\
& \equiv & f_{\rm wr}\cdot n_{\rm Ne,loc}
\label{eq:n_Ne,wr}
\end{eqnarray}
\begin{eqnarray}
n_{\rm Ne,loc,o} &=& \frac{\frac{\dot M_{\rm o}}{\mu _{\rm o}}(1-\cos \psi _{\rm o})}{\frac{\dot M_{\rm wr}}{\mu _{\rm wr}}(1-\cos \psi _{\rm wr})+\frac{\dot M_{\rm o}}{\mu _{\rm o}}(1-\cos \psi _{\rm o})}n_{\rm Ne,loc} \nonumber \\ 
& \equiv & f_{\rm o}\cdot n_{\rm Ne,loc}
\label{eq:n_Ne,o}
\end{eqnarray}
Here $\psi _{\rm wr}$ and $\psi _{\rm o}$ are the elevation angles of the Ne line-emission site viewed from the centers of the WR and O stars (see Fig.~\ref{fig:zahyou}).
These angles and the mixture fractions $f_{\rm wr}$ and $f_{\rm o}$ are listed in Table~\ref{tab:ring_val3}.

\begin{table*}
  \centering
  \caption{Values of the parameters used in the calculation to obtain the spatial extent of the Ne and O line-emission sites along the shock cone.  $\psi_{\rm wr/o}$ is the elevation angles of the Ne/O line-emission site viewed from the centers of the WR/O star (see Fig. \ref{fig:zahyou}). $f_{\rm wr/o}$ is the mixture fraction [see Equations~(\ref{eq:n_Ne,wr}) and (\ref{eq:n_Ne,o})].\label{tab:ring_val3}}
   {\tabcolsep = 8pt
  \begin{tabular}{lccccc} \hline
     & Phase & $\psi_{\rm wr}$ (degrees) & $\psi_{\rm o}$ (degrees)& $f_{\rm wr}$ & $f_{\rm o}$ 
     \\ \hline
     & K (0.816) & 17.5 & 95.2 & 0.30 & 0.70\\
     & A (0.912) & 16.2 & 87.4 & 0.29 & 0.71\\
    \ion{O}{vii, viii} 
     & L (0.935) & 16.9 & 91.8 & 0.29& 0.71 \\
     & B (0.968) & 16.8 & 93.1 & 0.28 & 0.72\\
     & D (0.987) & 13.7 & 77.0 & 0.27 & 0.73 \\ 
     \hline
     & K (0.816) & 17.1 & 92.4 & 0.29 & 0.71\\
     & A (0.912) & 15.7 & 84.2 & 0.29 & 0.71\\
    \ion{Ne}{ix, x} 
     & L (0.935) & 17.5 & 95.4 & 0.29 & 0.71 \\
     & B (0.968) & 16.5& 91.3& 0.28 & 0.72 \\
     & D (0.987) & 12.5& 68.3 & 0.27 &0.73 \\
    \hline
  \end{tabular}
  }
\end{table*}

In the same way as Equation~(\ref{eq:Neloc}), $n_{\rm e}$ and $n_{\rm Fe}$ are expressed as
\begin{eqnarray}
    n_{\rm e} &=& n_{\rm e,loc,wr} + n_{\rm e,loc,o} \label{eq:ne} \\
    n_{\rm Fe} &=& n_{\rm Fe,loc,wr} + n_{\rm Fe,loc,o} \label{eq:nFe}
\end{eqnarray}
For calculating $n_{\rm e,loc,wr}$ and $n_{\rm e,loc,o}$, we take the metal abundances of the WR wind from \citet[Table 3 of][]{2015PASJ...67..121S}:
\begin{equation}
\mbox{H} = 0, \mbox{ He} = 1, \mbox{ C} = 0.4, \mbox{ O} = 7.2\times 10^{-2}, \mbox{ Fe} = 4.16\times 10^{-4},
\label{eq:WRabundance}
\end{equation}
and those of the O star from \citet{1973asqu.book.....A} as
\begin{equation}
\mbox{H} = 11.7, \mbox{ He} = 1, \mbox{ C} = 3.9\times10^{-3}, \mbox{ O} = 7.8\times 10^{-3}, \mbox{ Fe} = 4.68\times 10^{-4}.
\label{eq:Oabundance}
\end{equation}
With these abundance sets, we can express $n_{\rm e,loc,wr}$ and $n_{\rm e,loc,o}$ with the particle number densities, and Equations~(\ref{eq:ne}) and (\ref{eq:nFe}) result in
\begin{eqnarray}
    n_{\rm e} &=& 0.77\,n_{\rm Ne,loc,wr} + 0.52\,n_{\rm Ne,loc,o} \label{eq:ne_2} \\
    n_{\rm Fe} &=& (6.5\,n_{\rm Ne,loc,wr} + 1.8\,n_{\rm Ne,loc,o})\,\times10^{-5} \label{eq:nFe_2}
\end{eqnarray}
By inserting Equations~(\ref{eq:n_Ne,wr}) and (\ref{eq:n_Ne,o}) into these two Equations, $n_{\rm e}$ and $n_{\rm Fe}$ can be expressed in terms of $n_{\rm Ne,loc}$ as
\begin{eqnarray}
    n_{\rm e} &=& (0.77\,f_{\rm wr} + 0.52\,f_{\rm o})\,n_{\rm Ne,loc} \label{eq:ne_3} \\
    n_{\rm Fe} &=& (6.5\,f_{\rm wr} + 1.8\,f_{\rm o})\times10^{-5}\,n_{\rm Ne,loc} \label{eq:nFe_3}
\end{eqnarray}
At phase K (0.816), $f_{\rm wr} = 0.30$ and $f_{\rm o} = 0.70$. Hence,
\begin{eqnarray}
    n_{\rm e} &=& 0.59\,n_{\rm Ne,loc} \label{eq:ne_4} \\
    n_{\rm Fe} &=& 3.2\times10^{-5}\,n_{\rm Ne,loc} \label{eq:nFe_4}
\end{eqnarray}

With $\Lambda_{\rm Fe,\odot} (T)$, $A_{\rm Fe,Allen}$, $n_{\rm e}$ and $n_{\rm Fe}$ being obtained here from Equations~(\ref{eq:ne_4}) and (\ref{eq:nFe_4}) and with $n_{\rm Ne,loc}$ from Equation~(\ref{eq:nNelineloc}), we can calculate $\varepsilon_{\rm Fe}(T)$ using Equation~(\ref{eq:varepsilon_Fe}); the results are summarised in Table~\ref{tab:ring_val1}. $\varepsilon_{\rm Fe}(T_{\rm Ne})$ at phase K (0.816) is $4.6\times 10^{-8}$ erg~cm$^{-3}$~s$^{-1}$.

As for $\varepsilon _{\rm brems}(T)$, we follow \citet{1979rpa..book.....R};
\begin{equation}
\varepsilon _{\rm brems}(T)=1.4\times 10^{-27}T^{1/2}n_{\rm e}\sum_{i}Z^2_in_i\bar g_{\rm B}\;\;[\mbox{erg cm$^{-3}$ s$^{-1}$}],
\label{eq:varepsilon_brems}
\end{equation}
where $\bar g$ is the Gaunt factor, which is set to 1.3.
In quite the same way as calculating $n_{\rm Fe}$ via Equations~(\ref{eq:nFe}) through (\ref{eq:Oabundance}), (\ref{eq:nFe_2}) and (\ref{eq:nFe_3}), 
we have calculated $\varepsilon _{\rm brems}(T)$ according to Equation~(\ref{eq:varepsilon_brems}), which are summarised in Table~\ref{tab:ring_val1}. 
At phase K (0.816), $\varepsilon_{\rm brems}(T)$ becomes $1.1\times10^{-8}$ erg~cm$^{-3}$~s$^{-1}$.

The same calculations can be done for O line-emission sites. Note that $f_{\rm wr}$ and $f_{\rm o}$ that appear in Equations~(\ref{eq:n_Ne,wr}) and (\ref{eq:n_Ne,o}) are different from those of the Ne line-emission sites.

\bsp
\label{lastpage}
\end{document}